\documentclass[12pt, a4paper]{article}
\pdfoutput=1

\usepackage{amsmath}
\usepackage{amsfonts}
\usepackage{amssymb}
\usepackage{graphicx, rotating}
\usepackage{epstopdf}
\usepackage{epsfig}
\usepackage{latexsym}
\usepackage{graphicx}
\usepackage{color}
\usepackage{amsmath,amssymb}
\usepackage{cite}
\usepackage{slashed}
\usepackage{hyperref}

\numberwithin{equation}{section}

\hypersetup{colorlinks, citecolor=bluscuro, linkcolor=bluscuro, urlcolor=bluscuro}
\definecolor{rossos}{rgb}{0.8,0.2,0.3}
\definecolor{bluscuro}{rgb}{0.15, 0.2, 0.9}

\makeatother   

\newcommand{\bb}[1]{\textbf{#1}}
\newcommand{\vv}{ ``}

 \def\be   {\begin{equation}}   \def\ee   {\end{equation}}
 \def\ba   {\begin{array}}      \def\ea   {\end{array}}
 \def\bea  {\begin{eqnarray}}   \def\eea  {\end{eqnarray}}
 \def\bean {\begin{eqnarray*}}  \def\eean {\end{eqnarray*}}

\begin{document}


\vspace{0.5cm}
\begin{center}

\def\thefootnote{\fnsymbol{footnote}}

{\Large \bf Restructuring the Italian NHS: \\
  \vskip 0.3cm a case study of the regional hospital network}
\\[1.2cm]
{\large \textsc{Carlo Castellana}} 
\\[1cm]

{\small \textit{ Management School in Clinical Engineering,\\
University of Trieste, Italy}}

\vspace{.2cm}

\end{center}

\vspace{.8cm}

\begin{center}
\textbf{Abstract}
\end{center}

\noindent
One of the main issues affecting the Italian NHS is the healthcare deficit: according to current agreements between the Italian State and its Regions  public funding of regional NHS is now limited to the amount of regional deficit and is subject to previous assessment of strict adherence to constraint on regional healthcare balance sheet. Many Regions with previously uncontrolled healthcare deficit have now to plan their \vv Piano di Rientro" (PdR) and submit it for the approval of the Italian Ministry of Economy and Finance. 
Those Regions that will fail to comply to deficit constraints will suffer cuts on their public NHS financing.
A smart Health Planning can make sure health spending is managed appropriately. Indeed a restructuring of the Italian healthcare system has recently been enforced in order to cope for the clumsy regional healthcare balance sheets.
Half of total Italian healthcare expenditure is accounted by hospital services which therefore configure as one of the main restructuring targets.
This paper provides a general framework for planning a re-engineering of a hospital network. This framework is made of economic, legal and healthcare constraints. We apply the general framework to the particular case of Puglia region and explore a set of re-engineered solutions which to different extent could help solve the difficult dilemma: cutting costs without worsening the delivery of public healthcare services.

\def\thefootnote{\arabic{footnote}}
\setcounter{footnote}{0}
\pagestyle{empty}

\newpage
\pagestyle{plain}
\setcounter{page}{1}

\section{Introduction} \label{Introduction}
\noindent
One of the main issues which is currently threatening the public finances of the Italian Republic is its huge amount of healthcare deficit. 
The Italian public healthcare expenditure accounts for 9.5\% of GDP \footnote{http://www.oecd.org}.
Italy's healthcare system is a regionally based National Health Service (NHS) that provides universal coverage free of charge at the point of service. 
There are two types of healthcare financing: public and private. With a ratio between public to private financing of 80:20 \footnote{Source: OECD, 2008}, the Italian NHS can be classified as a publicly financed system.

There is considerable variation between the North and the South in the quality of healthcare facilities and services provided to the population, with significant cross-regional patient flows, particularly to receive high-level care in tertiary hospitals.
The national level is responsible for ensuring the general objectives and fundamental principles of the national healthcare system. Regional governments, through the regional health departments, are responsible for ensuring the delivery of a benefits package through a network of population-based health management organizations \footnote{ASLs or Azienda Sanitaria Locale, i.e. local health enterprises.} and public and private accredited hospitals.
The health budget is determined centrally and financed partly by employers and employees contributions with the Government paying the balance directly.
According to current agreements between the Italian State and its Regions \cite{ConferenzaStatoRegioni} public funding of regional NHS is now limited to the amount of regional deficit and is subject to previous assessment of strict adherence to constraint on regional healthcare balance-sheet. Many Regions with previously uncontrolled healthcare deficit have now to plan their \vv Piano di Rientro" (PdR) and submit it for the approval of the Italian Ministry of Economy and Finance. 
Those Regions that will fail to comply to deficit constraints will suffer cuts on their public NHS financing.

A restructuring of the Italian healthcare system has recently been enforced by Italian Public Authorities in order to cope for the clumsy regional healthcare balance-sheets. 
In a previous paper \cite{Castellana} we showed that the estimated impact of the current economic crisis on Italian public healthcare expenditure is comparable to the healthcare deficit of Italian Regions, meaning that it could seriously worsen an already difficult situation. Henceforth it is essential at this stage that policies of public health planning face the difficult problem of cutting costs without reducing healthcare services. A smart re-organization of the NHS could make sure health-spending is managed appropriately.
Half of total Italian healthcare expenditure is accounted by hospital services \footnote{Source: 2008 data, Sistema informativo Sanitario, Ministero della Salute}. This means that it is very likely that a main portion of the regional healthcare deficit should be attributed to inefficiencies at the hospital level which, as a consequence, becomes one of the main restructuring target of Public Authorities in charge of health planning.

By focusing on Puglia \footnote{\textit{Puglia} is the Italian name, while \textit{Apulian} is the adjective.}, one of the 20 Italian Regions, in this paper we provide a general framework for planning a re-engineering of a hospital networkwork. This framework is made of economic, legal and health-care constraints. 
The general framework is applied to the particular case of the hospital networkwork of Puglia and we provide a set of re-engineered solutions which to different extent could help achieving the aforementioned goal of cutting costs without worsening the delivery of public healthcare services.

We will start by listing the legal constraints of the problem, then we will overview the methodology. The last two sections will present the results and will show the limits and perspectives of our proposal.

\noindent 
There are several reasons for having selected the Puglia case-study. Just to mention a few of them:
\begin{itemize}
\item  Puglia is one of those Regions having a relevant negative healthcare deficit  \footnote{EUR 300 mln, according to \textit{Ragioneria Generale dello Stato, Ministero dell'Economia e delle Finanze}, 2009}. By the end of 2010 Puglia has submitted its \vv Piano di Rientro" (PdR) for the approval of the Ministry of Economy and Finance \cite{PdR};
\item  while for other Regions \footnote{E.g. Molise, Campania, Lazio.} the PdR was quickly approved by the Italian Ministry of Economy and Finance, the Apulian PdR had a long route before getting approved. The Apulian healthcare system did not seem to fit easily within current national standards up to the extent that the approved PdR requested an impending 2.200 bed-cuts and closedown of 18 hospitals which caused a heated public debate;
\end{itemize}

\section{The Apulian healthcare system}
\label{sec: intropu}
%
\noindent 
The first step that we have to make in order to re-engineer the Apulian hospital networkwork is to have a clear picture of the system we are going to restructure and of the main constraints that we have to comply with. This introductory section reports some data on the organization of the Apulian RHS and on the legal constraints that we will have to face in our task.

\subsection{Statistics} \label{sec: constraints}
\begin{figure}[!htbp]
\centering
\includegraphics[scale=0.5]{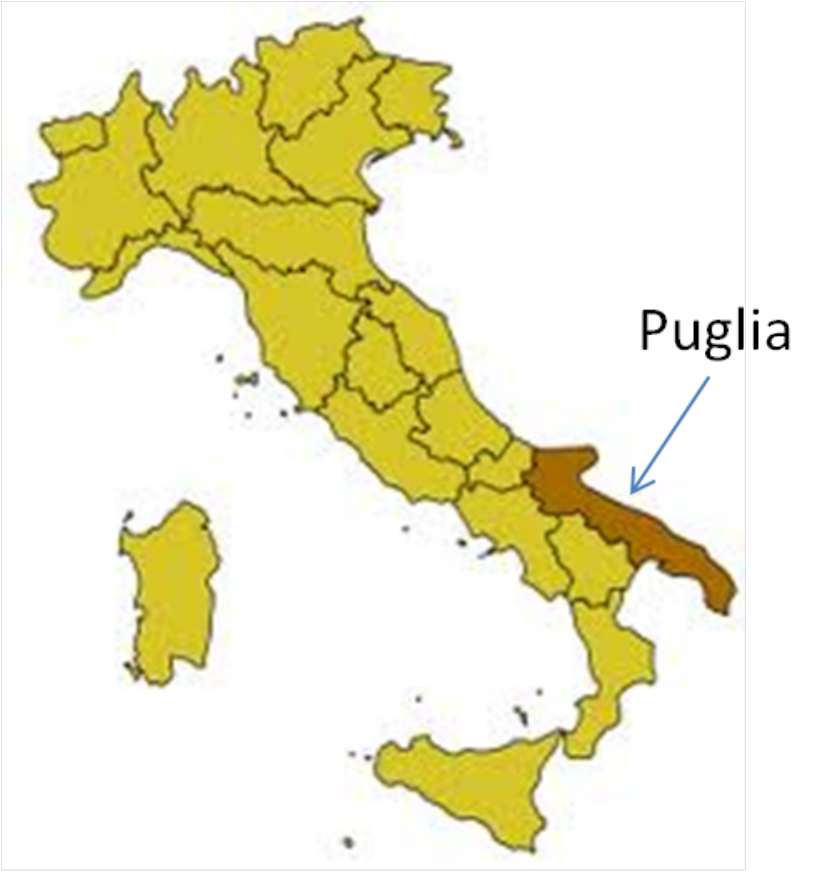}
\caption{Puglia is a Region located in south-east Italy.}
\label{fig: Osp_2}
\end{figure}

\begin{figure}[!htbp]
\centering
\includegraphics[scale=1.2]{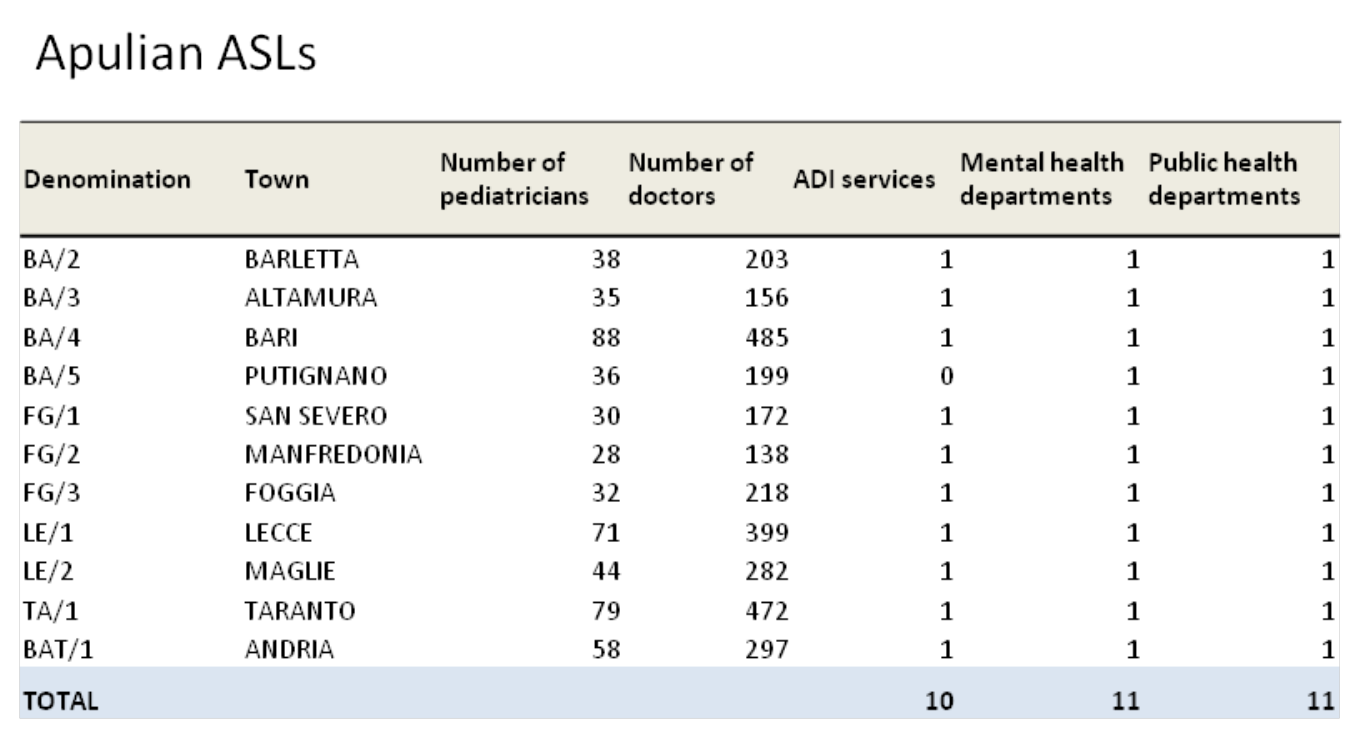}
\caption{As in any other Italian Region, the Apulian regional public healthcare service is organized in ASL. \textit{Source: Dati ASL 2007, Ministero della Salute}. }
\label{fig: Osp_5}
\end{figure}

Puglia is a Region of Southern Italy (see fig. \ref{fig: Osp_2}). 
\noindent With its 4.076.546 residents \footnote{Source: ISTAT, 2008} (6,8\% of total Italian population, 2008) it is one of the most populated Italian Regions and contributes significally to total NHS expenditure. Data as of 2008 report a healthcare expenditure of EUR 7 bn \footnote{About 7\% of Italian NHS expenditure. Source: Sistema informativo sanitario, \href{www.salute.gov.it}, data as of 2008.}. Even if not efficient as the one of many northern Italian Regions, its healthcare system represents a reference point for populations of its two surrounding Regions, Basilicata and Molise \footnote{Passive healthcare mobility toward Puglia: Basilicata 10\%, Molise 2\%.}.
The Apulian regional Healthcare System (RHS) is organized within 11 ASLs (fig. \ref{fig: Osp_5}).

\begin{figure}[!htbp]
\centering
\includegraphics[scale=0.95]{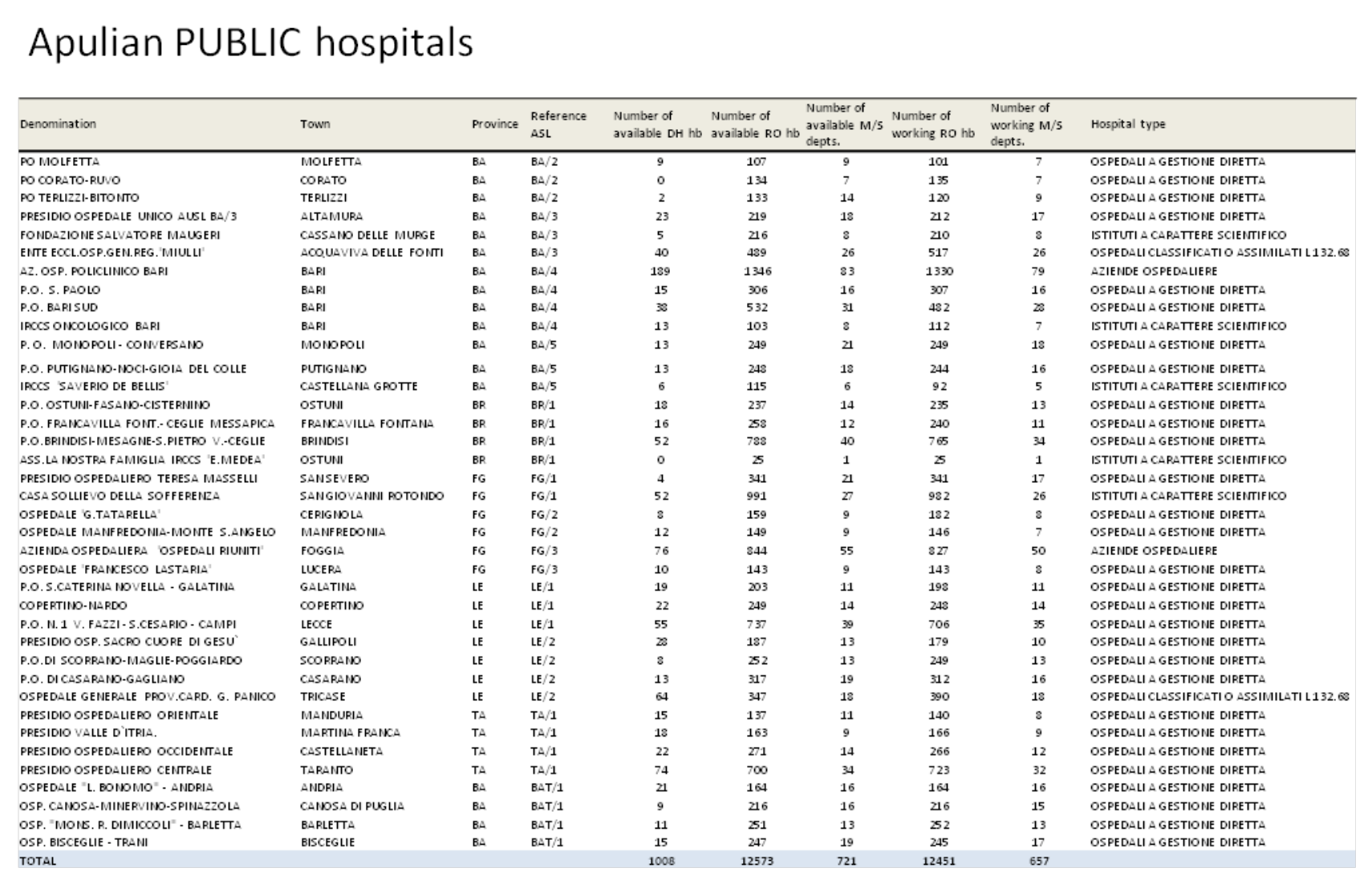}
\caption{Public hospitals delivering healthcare services in Puglia. Additional information on hospital beds is available (e.g. split between RO and DH beds, allocation among different Specialties). hb: hospital beds; M/S depts: Medical/Surgical internal departments (e.g. Cardiology, General Surgery, etc.) \textit{Source: Dati SDO 2006, Ministero della Salute}. }
\label{fig: Osp_4}
\end{figure}

\begin{figure}[!htbp]
\centering
\includegraphics[scale=1.1]{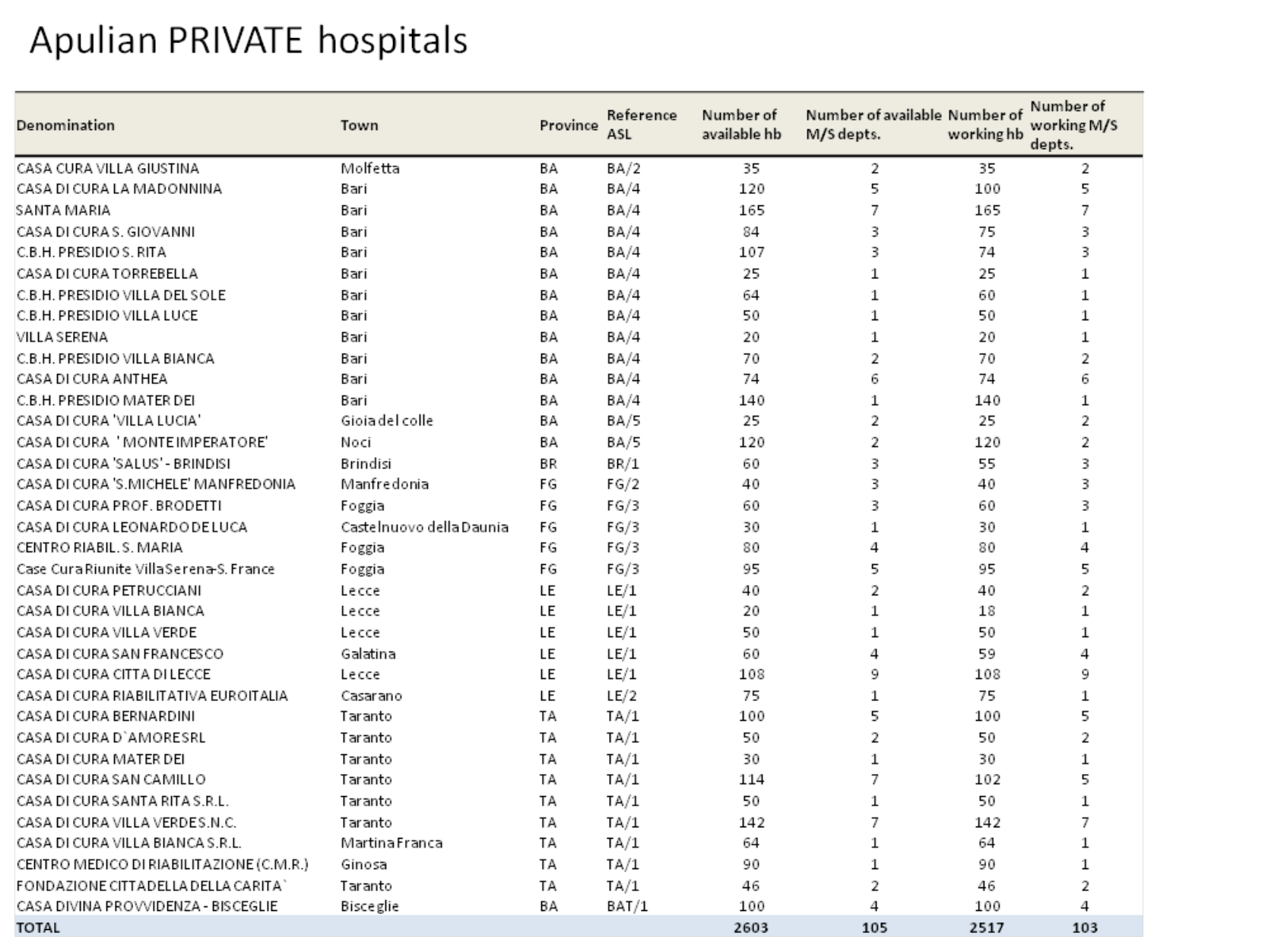}
\caption{Private hospitals delivering healthcare services in Puglia. Unfortunately no additional information on hospital beds is available (e.g. split between RO and DH beds, allocation among different Specialties). hb: hospital beds; M/S depts: Medical/Surgical internal departments (e.g. Cardiology, General Surgery, etc.) \textit{Source: Dati SDO 2006, Ministero della Salute}. }
\label{fig: Osp_3}
\end{figure}

The Apulian hospital networkwork counts 38 public hospitals (1 Policlinico Universitario and 3 IRCCS\footnote{Istituto di Ricovero e Cura a Carattere Scientifico.}.) and 36 private NHS-accredited hospitals (see fig. \ref{fig: Osp_4} and \ref{fig: Osp_3}). The total number of hospital beds is around 13.000 in public hospitals and 2.500 in private hospitals \footnote{Source: Sistema informativo sanitario, \href{www.salute.gov.it}{[www.salute.gov.it]}, data as of 2007-08.}. \\

Most of the data used in this chapter and related to Apulian RHS have been taken from the Italian Ministry of Health public database. This was the most reliable data-source to which we had access.
Unfortunately, the information available lacks of completeness. First of all, much more details are available for public hospitals than for private ones. For \bb{public} hospitals the total number of DH hb \footnote{\textit{hb} stands for \vv hospital beds". \textit{DH} and \textit{RO} respectively stand for \vv day hospital" and \vv ordinary" admissions.} and the total number of RO hb are available. Moreover, the split between DH and RO hb among the different specialties (e.g. Internal Medicine, Cardiology, etc) for each single public hospital is available. 

For \bb{private} hospitals the Ministry of Health reports the \vv total number of available/working beds" only, without clarifying whether that number refers to RO hb only or if it includes DH hb as well. Moreover, for private hospitals details on how that number of beds is split among different specialty departments within the same entity have not been provided. 

In addition to that, no information is provided in relation to each medical/surgical department being dedicated to acute or rehabilitation cares (e.g. in fig. \ref{fig: Osp_3}). 

\noindent Let's make an example (see fig. \ref{fig: Osp_4} and \ref{fig: Osp_3}).\\ 
For a public hospital as \vv P.O. Putignano-Noci-Gioia del Colle" we know: (i) total number of hb: 13 DH hb and 248 RO hb, (ii) total number of specialty departments: 18, (iii) specialties: general medicine, cardiology, general surgery, etc., (iv) hb allocation among specialties: general medicine has 2 DH hb and 36 RO hb.\\
For a private hospital as \vv Santa Maria" we know: (i) total number of hb: 165 bd, (ii) total number of specialty departments: 7. No additional information is available.

This will represent one of the main limits of our restructuring plan. Indeed without information on the current hb allocation within the private sector (which represents 15 - 20\% of total number of hb) we will not be able to assess the re-engineered distribution of hb neither in terms of public/private splitting, nor in terms of allocation to each single specialty.\\

\begin{figure}[!htbp]
\centering
\includegraphics[scale=1.0]{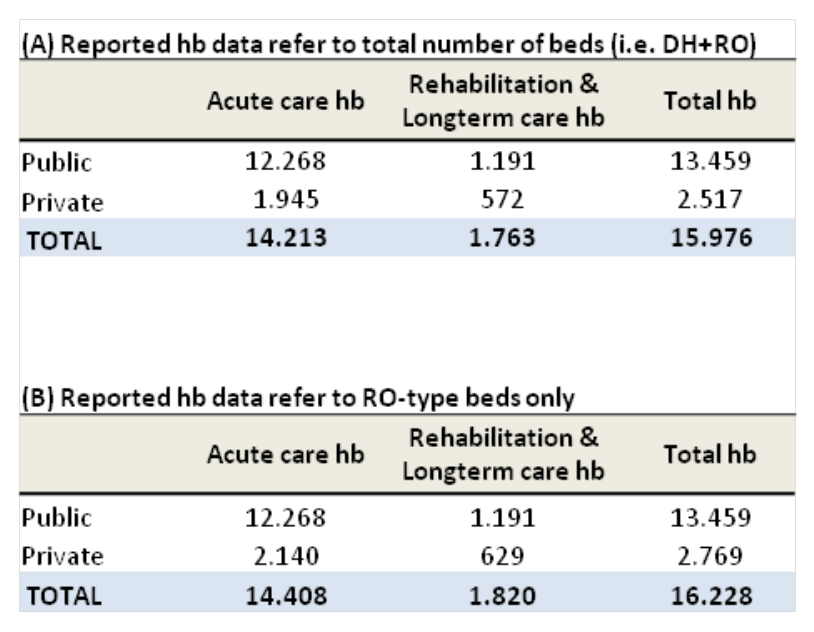}
\caption{The data available on the Ministry of Health database include: (\textit{public structures}) (i) number of acute DH and RO hb, (ii) number of rehabilitation and longterm care hb; (\textit{private structures}) (iii) number of acute hb, (iv) number of rehabilitation and longterm care hb. For (iii) and (iv) original source does not specify whether it refer to the sum of DH+RO hb or to RO hb only. Two assumptions are then made: (A) upper table, original source refers to the sum of DH+RO hb; (B) lower table, original source refers to the RO hb only. \textit{Data NHS 2007-08, Ministero della Salute}.}
\label{fig: Osp_6}
\end{figure}

Given that it was not specified whether or not the reported numbers of rehabilitation and longterm care hb (for public hospitals) and of both acute hb and rehabilitation and longterm care hb (for private hospitals) included the number of DH hb, we took into account both alternatives. Both alternatives are reported in fig. \ref{fig: Osp_6}.
The results of the two alternatives differed quantitatvely but not qualitatively. In the following we decided to report the calculations performed for the (A) alternative of fig. \ref{fig: Osp_6} only.

\subsection{Legal constraints} \label{sec: constraints}
There are some strict constraints that must be met when planning for a hospital networkwork. The importance of complying to such constraints stands from the restrictions imposed by the Italian Government: compliance to those constraints is required for each Region to benefit of additional State financing to regional NHS.
A list of the main constraints follows:
\begin{enumerate}
\item  \bb{Hospital beds} \footnote{Intesa Stato-Regioni of 3 Dec 2009. Additional details can be found on art. 6 of the original document.}: the total \bb{number of hospital beds} (hb) should \bb{not exceed 4 hb every 1.000 residents}. This total rate applies to the sum of the five type of admissions: acute ordinary (ARO), acute day hospital (ADH), rehabiltitation ordinary (RRO), rehabilitation day hospital (RDH) and longterm care (LTC). Additional constraints apply to total acute admissions and total rehabilitation admissions: 
	\begin{enumerate}
	\item the total \bb{number of beds} \footnote{i.e. ARO and ADH.} for \bb{acute} admission should \bb{not exceed 3,3 hb for 1.000 residents};
	\item the total \bb{number of beds} \footnote{i.e. RRO, RDH and LTC.} for \bb{rehabilitation and longterm care} admission should \bb{not exceed 0,7 hb for 1.000 residents}.
	\end{enumerate}

\item  \bb{Hospitalization rate} \footnote{Intesa Stato-Regioni of 23 Mar 2005.}: the total hospitalization rate \footnote{The agreement refers to the crude rate.} should \bb{not exceed 180 persons every 1.000 residents}. This total rate applies to the sum of the five type of admissions (ARO, ADH, RRO, RDH and LTC).

\item  \bb{Percentage of Day Hospital admissions} \footnote{Intesa Stato-Regioni of 23 Mar 2005.}: the \bb{number of total day hospital admissions} \footnote{i.e. ADH and RDH}, calculated as the ratio between the number of day hospital admissions and the total number of admissions, {should not fall below 20\%}.

\item  \bb{Percentage of Day Hospital beds} \footnote{L. n. 662 of 23 Dec 1996, art. 1.}: the \bb{number of day hospital beds} \footnote{i.e. ARO and RDH}, calculated as the ratio between the number of day hospital beds and the number of total beds, should \bb{not fall below 10\%}.
\end{enumerate}

\begin{figure}[!htbp]
\centering
\includegraphics[scale=1.1]{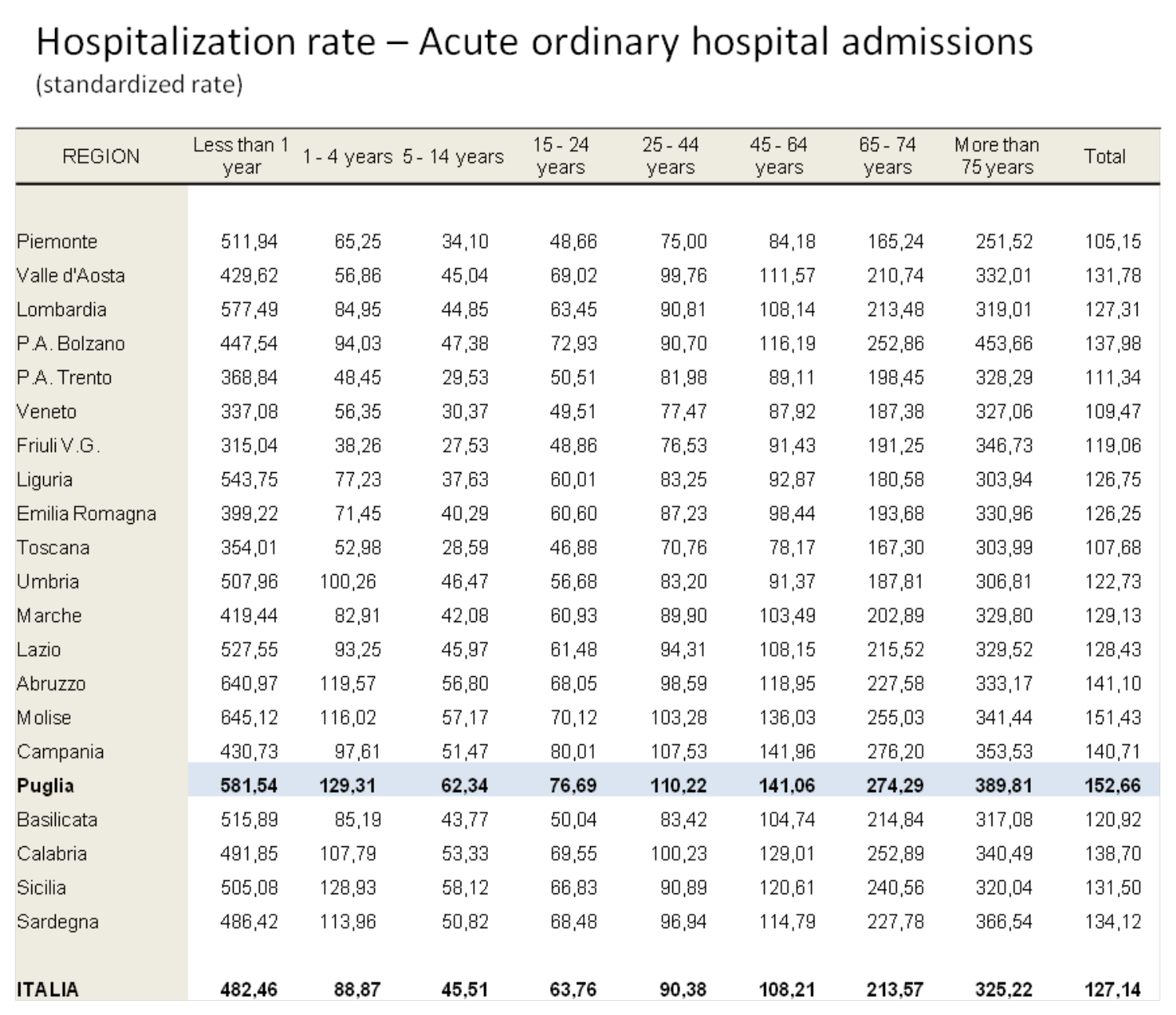}
\caption{Hospitalization rate among Italian Regions. The rate refer to acute ordinary admissions only (i.e. it does not include: acute day hospital admissions, rehabilitation ordinary and day hospital admissions and longterm care admissions. The rate is shown as number of admissions for 1.000 inahbitants. The rate has been standardized on the Italian population (i.e. it is not a crude rate). \textit{Dati SDO 2008, Ministero della Salute}.}
\label{fig: Osp_1}
\end{figure}

The Apulian hospitalization rates are the highest among Italian Regions: the 2008 standardized hospitalization rate for acute ordinary admissions (ARO, see fig. \ref{fig: Osp_1}) was 152,66 for Puglia versus a national average of 127,14 \footnote{Data publicly available at \href{www.salute.gov.it}{[www.salute.gov.it]}, Sistema informativo sanitario.}. Obviousely this data is not comparable with the above legal constraint of 180 since this last refers to the five type of admissions (while 152,66 refers to the ARO only), but anyway gives us an idea of why the Apulian case seems to be more \vv difficult to treat" if compared to other cases.

\noindent Based on 2008 data, the density of hospital beds does not seem to fall above the 4,0 threshold.

\subsection{Apulian hospitalization needs} \label{sec: hospitneeds}
After a quick overview of the Apulian RHS, the second step is quantifying the hospitalization needs of Apulian population. This is clearly a crucial point of our planning since obviuosly the restructuring of the hospital network will have to comply to the most fundamental constraint of guaranteeing an appropriate delivery of healthcare services according to LEAs \footnote{The fundamental levels of healthcare assistance guaranteed by the Italian NHS \cite{LEA}.} which is the reason of being of the Italian NHS itself. 

It is worth mentioning that, even if current Apulian RHS suffers from huge and continuing deficits, the solution to the problem cannot be financial only. A good planning will size hospital beds in compliance to:
\begin{enumerate}
\item 	\bb{healthcare constraints}: the hospital network should be sized in relation to the specific healthcare needs of Apulian population;
\item 	\bb{legal constraints}: we have to make sure that all legal constraints of section \ref{sec: constraints} are satisfied;
\item 	\bb{financial constraints}: the final goal is to make the hospital network to work more efficiently, such that resources are better employed and healthcare deficit gets reduced. The re-engineering plan should generate financial value.
\end{enumerate}
The next question is thus: what are the healthcare constraints we have to comply with? I.e. how can we quantify the healthcare needs of the Apulian population?\\

\begin{figure}[!htbp]
\centering
\includegraphics[scale=0.62, angle=0]{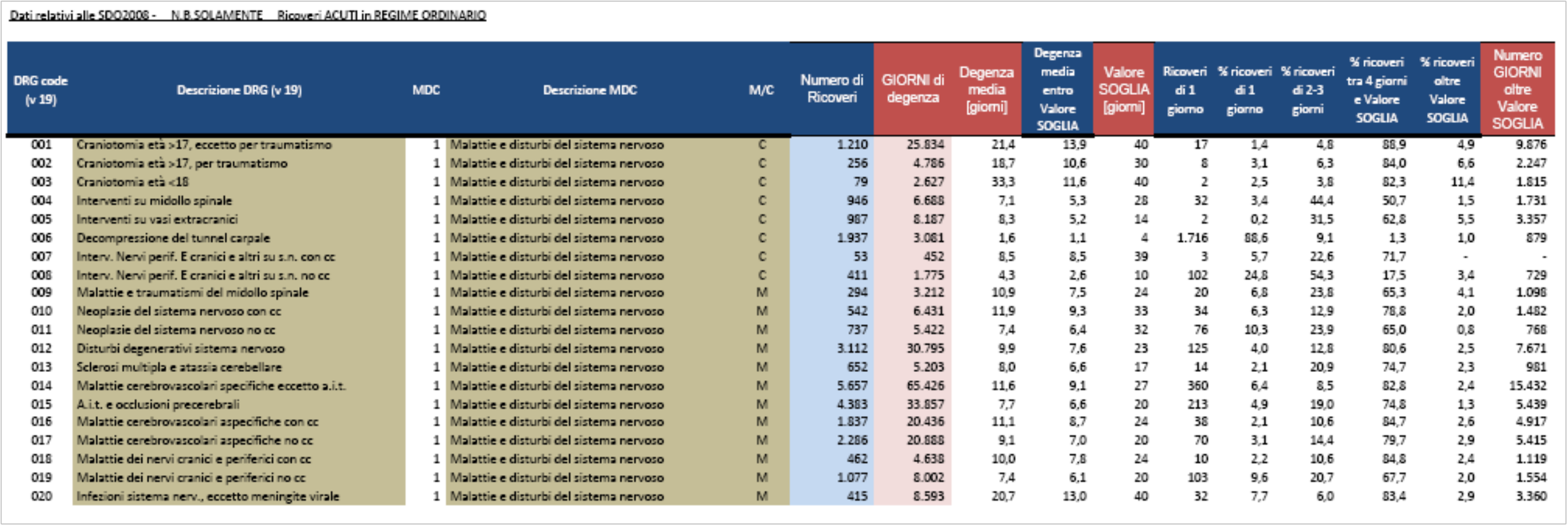}
\caption{List of DRGs calculated on 2008 Apulian hospital acute admissions. The original list includes 513 rows: we report the first 66 rows only. For each DRG various informations are reported (from left to right): medical or surgical DRG (M/C), number of admissions, number of hospitalization days, average number of hospitalization days, average number of hospitalization days below threshold, threshold on hospitalization days for single admission, number of 1 day admissions, percentage number of 1 day admissions (and of 2-3 days admissions, and of admissions between 4 days and threshold), percentage number of admissions with an above threshold duration, total number of hospitalizaion days above threshold. \textit{Dati SDO 2008, Ministero della Salute}.}
\label{fig: Osp_7}
\end{figure}

Statistical data on Apulian hospitalization are summarized in the SDO 2008 report of the Italian Ministry of Health \cite{SDO2008}. As it can be seen from fig. \ref{fig: Osp_7} - where we report few rows only of the original data - the informations available are exacly what we need: number of hospital admissions, total lenght of hospital stays, number of potentially inappropriate admissions (i.e. both those with a 1 day duration and those with a duration above threshold). The above information should be intended as the Apulian \vv demand" for acute ordinary hospitalizations.

\subsection{LEAs with high risk of inappropriateness}
\begin{figure}[!htbp]
\centering
\includegraphics[scale=0.5]{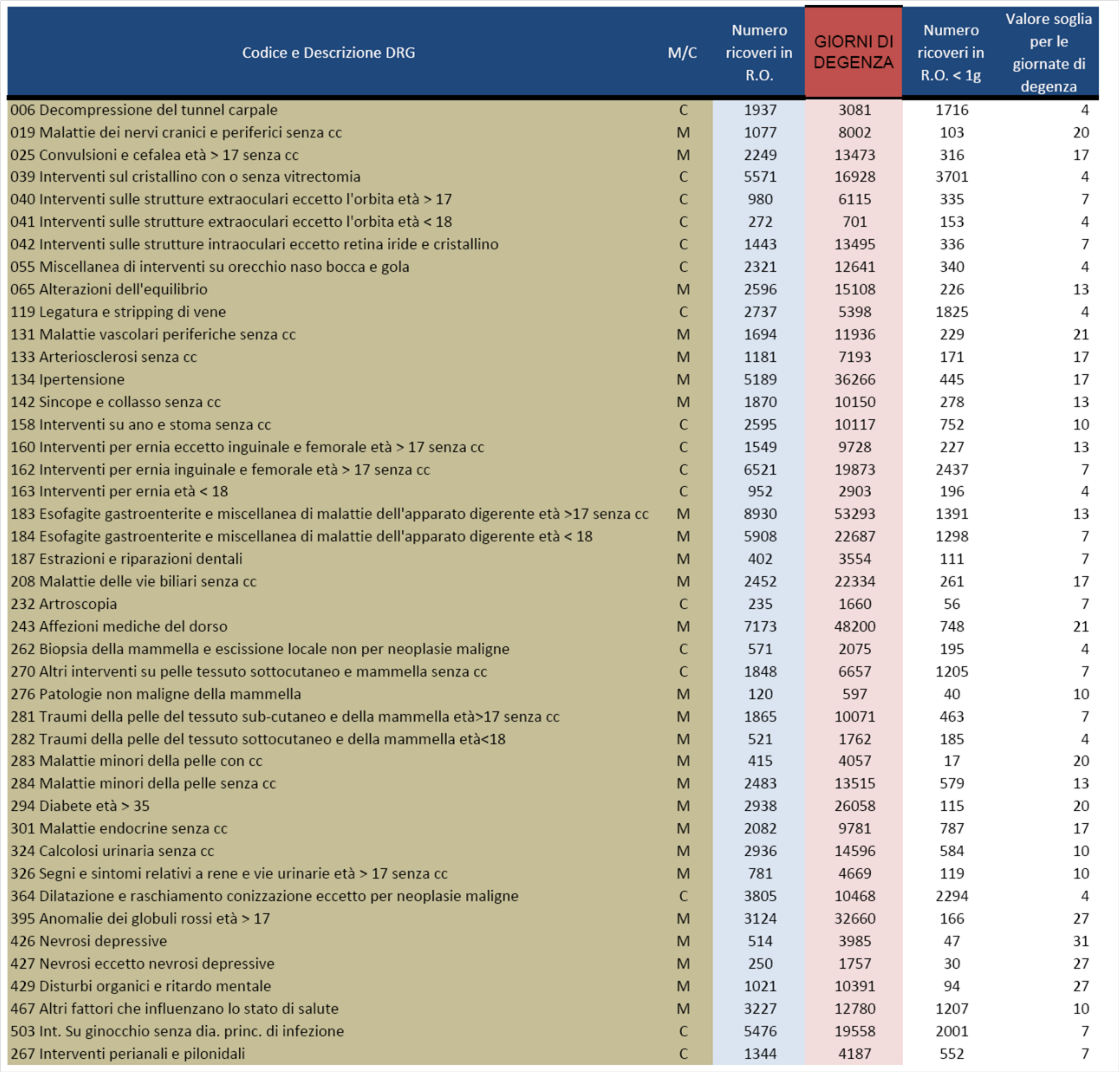}
\caption{List of DRGs with a high risk of inappropriateness according to D.P.C.M. 29 Nov 2001. \textit{Dati SDO 2008, Ministero della Salute}.}
\label{fig: Osp_8}
\end{figure}

\begin{figure}[!htbp]
\centering
\includegraphics[scale=0.45]{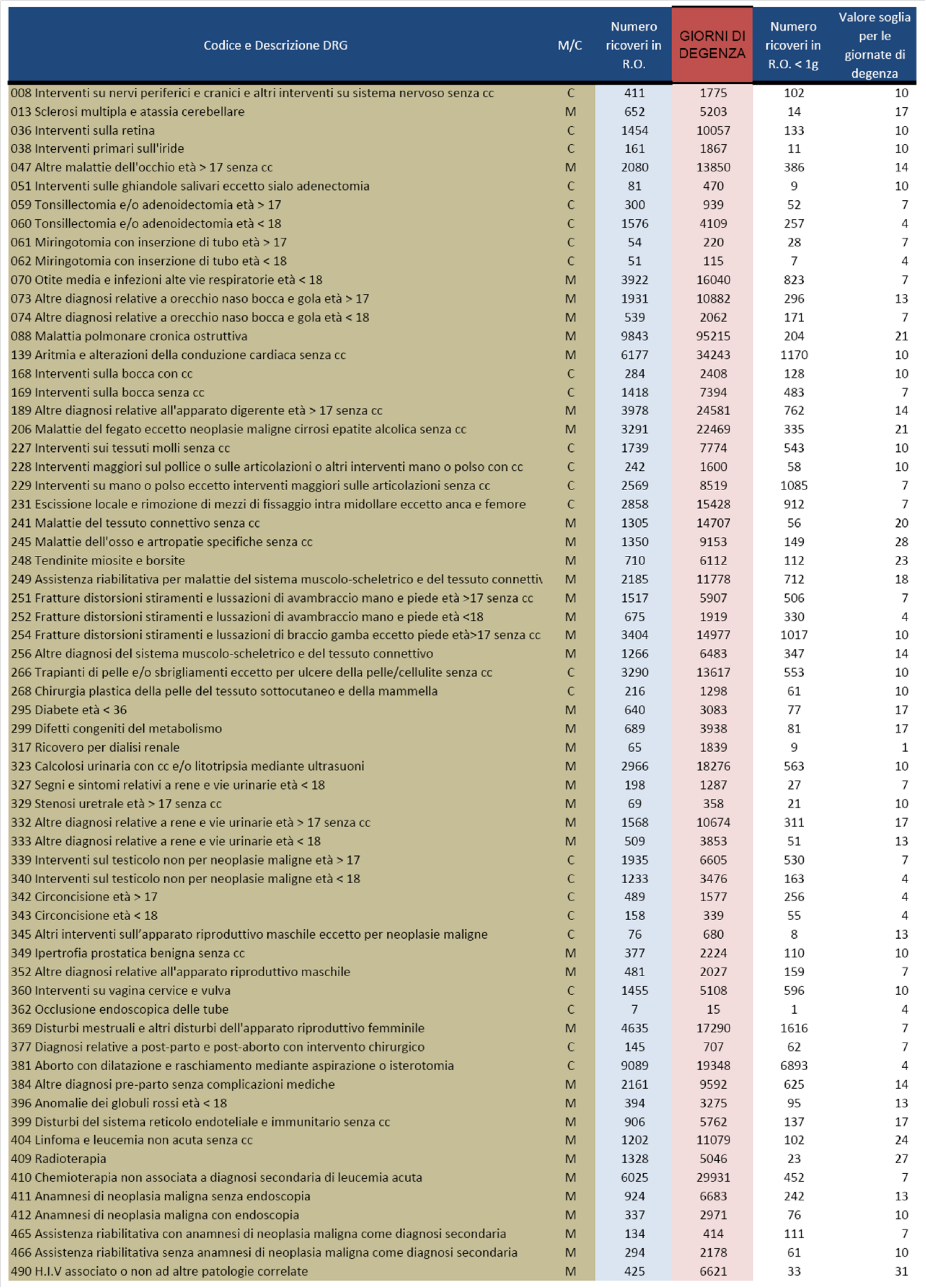}
\caption{List of \bb{additional} DRGs with a high risk of inappropriateness according to \vv Patto per la Salute" 2010-2012. \textit{Dati SDO 2008, Ministero della Salute}.}
\label{fig: Osp_9}
\end{figure}

The D.P.C.M. 29 Nov 2001 \cite{DPCM29Nov11} makes a list of DRGs qualified as \vv with a high risk of inappropriateness". DRGs that fall within that list should appropriately be delivered as day hospital or ambulatory services. Nine years have passed since 2001 and during the last few years the Regions are adopting their own regional laws that enforce the delivery of those DRGs in day hospital or ambulatory services.
The list of 43 DRGs is reported in fig. \ref{fig: Osp_8}. For each DRG additional informations related to acute hospital admissions of Apulian residents are reported.
\\
The \vv Patto per la Salute 2010-2012" has added other 65 DRGs (fig. \ref{fig: Osp_9}).

When planning for the restructured hospital network we have to make sure these DRGs are treated appropriately: if these should appropriately be delivered as day hospital or ambulatory services, then our aim will be to transfer as much admissions as possible from acute ordinary (ARO) to acute day hospital (ADH) and ambulatory services (AMBUL). 
The way those 108 DRGs are being accounted is becoming standardized among the various Regions. For example, the plan for the hospital network of Regione Basilicata has been made without taking into account the admissions with those 108 DRGs \cite{PdRBasilicata}.
A less aggressive approach has been taken by Regione Molise that in its \vv Piano di Rientro" \cite{PdRMolise} plans its hospital network based on a transfer of 85\% of the total days of hospital stays associated to those DRG from ordinary admission to day hospital.
A more conservative approach is suggested by Regione Campania where in its \vv Riassetto della rete ospedaliera e territoriale" \cite{PdRCampania} trasfers from RO to DH or AMBUL around 50-70\% of ordinary admissions.



\section{Methods}
\noindent
\subsection{Fundamentals} \label{sec: fundamentals}
The calculations we have performed are based on a simple relationship between supply and demand of hospitalization services. As basic economics teaches, in an equilibrium state the total demand of hospitalization services should be equal to total supply. How can we quantify \vv a certain amount" of hospitalization services? A possibility would be in terms of hospitalization days. At the end of the day, patients, through their doctors, translate their health care needs into admissions to hospitals.
Within this framework we can write:
\begin{eqnarray}
\textit{Number of days}_{Demand} &=&  d_{M} \cdot \alpha \cdot P = d_{M} \cdot R \nonumber \\
\textit{Number of days}_{Supply}  & = & \beta \cdot NB_{RO} \cdot 365 = \beta \cdot n  \cdot 365 \cdot P 
\end{eqnarray}
where $NB$ is the number of hospital beds, $P$ is the population, $d_{M}$ is the average duration of hospital stays, $R$ is the number of admissions, $n$ is the density of hospital beds \footnote{Number of beds every 1.000 residents.} and $NB$ is the absolute number of beds within the hospital network.
The $\alpha$ parameter is the hospitalization rate, while the $\beta$ is the bed utilization rate, a measure of bed efficiency.
Under assuptions of equilibrium, we have that the demand equals the supply and hence:
\begin{equation} \label{eq: fundam}
d_{M} \cdot \alpha =  \beta \cdot n  \cdot 365
\end{equation}
which is the fundamental relationship that will be used to plan the restructured net.

The above relationship, while being correct for ordinary admissions, needs some adjustment when day hospital admissions are taken into account. A DH bed usually serves two patients ($A = 2$) during the same day (at least this should be the target bed turnover). A single DH service is usually delivered in two patient accesses ($acc = 2$) \footnote{See D.G.R. Lazio 423/05 and the restructuring plan of hospital services of Regione Lazio.}, which normally happens in two different days. By taking into account that generally day hospital services run 5 days a week, in order to have a $\beta$ that is related to the intrinsic utilization of DH services an annual 250 day-service should be assumed:
\begin{eqnarray}
\textit{Number of accesses}_{ Demand} &=& \textit{acc}  \cdot R  \nonumber \\
\textit{Number of accesses}_{Supply}&=& \beta \cdot A \cdot NB_{DH} \cdot 250 
\end{eqnarray}
where $NB_{DH}$ is the absolute number of DH beds.

\noindent By putting together the above two equations we have the general formula when both RO and DH admissions are taken into account:
\begin{equation} 
\beta \cdot NB= \frac{D_{DH}}{A \cdot 250} +  \frac{D_{RO}}{365}
\end{equation}
where $NB = NB_{DH} + NB_{RO}$ is the total absolute number of hospital beds, $D_{DH}$ is the total number of annual DH accesses and $D_{RO}$ is the total number of RO hospitalization days.

\subsubsection{Financial impact}
In the next section we will present some simulated scenarios of the restructured hospital network where, other than showing how resources can be rearrangements to comply with all aforementioned constraints, an indication of the financial impact is provided.
Some important notes to understand the real meaning of those numbers will be delivered in sec. \ref{sec: limitsper}. Within this subsection we anticipate that calculating the \textit{total} financial profit and loss (P\&L) of the real restructuring depends on the real implementation of the re-engineering plan. 
The only financial P\&L that will be calculated is that coming from the simple change in number and type of hospital beds.

The methodology we have used to calculate the financial impact of the re-engineering is based on the following cost structure \footnote{Based on \cite{PdRCampania} and on private interviews with three current Apulian hospital managers.}:
\begin{itemize}
\item 	acute hospital bed: EUR 250.000 p.a. \footnote{per annum};
\item    rehabilitation and longterm care hospital bed: about 65\% of the acute hospital bed cost, i.e. EUR 162.500 p.a.;
\item    residential and semiresidential services (RSA): about 30\% of the acute hospital bed cost, i.e. EUR 75.000 p.a.; 
\item 	ambulatory services: a reasonable estimate of the cost of an average single ambulatory service to be around EUR 200 \footnote{We assumed an opening time of 8 hours a day. The estimate takes into account the variability among specialist services (e.g. a CT costs much more than a simple general visit). Costs have been estimated based on \cite{TariffeAmbul}.}.
\end{itemize}

\subsection{Methodology}
\begin{figure}[!htbp]
\centering
\includegraphics[scale=0.95]{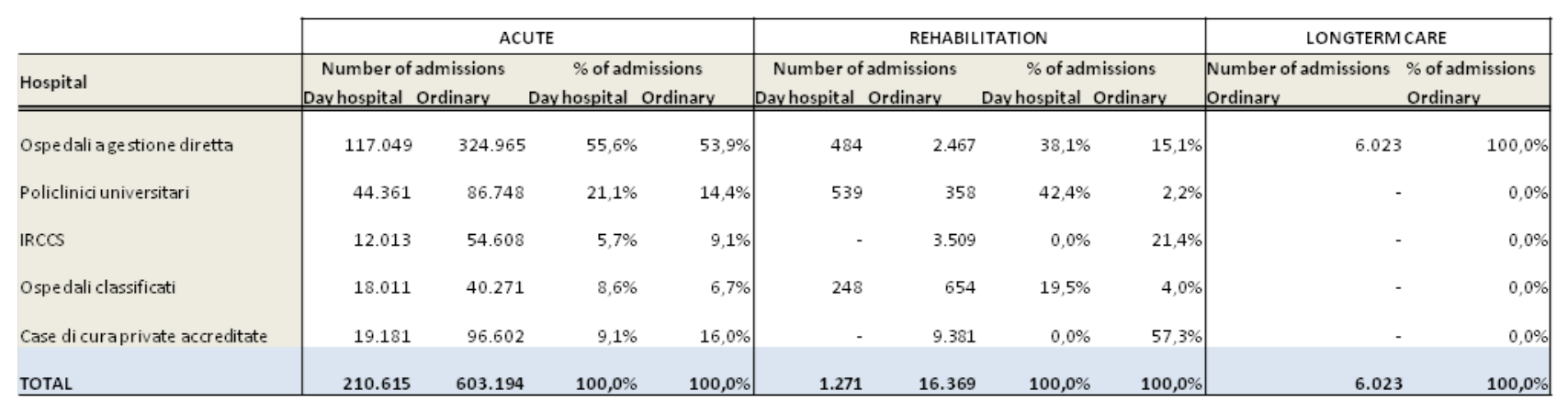}
\caption{Distribution of hospital admissions among different hospital structures. Note: private hospitals reported zero admissions to DH rehabilitation services. Even if we have not been able to confirm that no DH beds are available for rehabilitation, unless data reported on SDO 2008 are wrong, based on this data we can deduct that there are no DH private beds for rehabilitation. \textit{Dati SDO 2008, Ministero della Salute}.}
\label{fig: Osp_11}
\end{figure}

\begin{figure}[!htbp]
\centering
\includegraphics[scale=0.95]{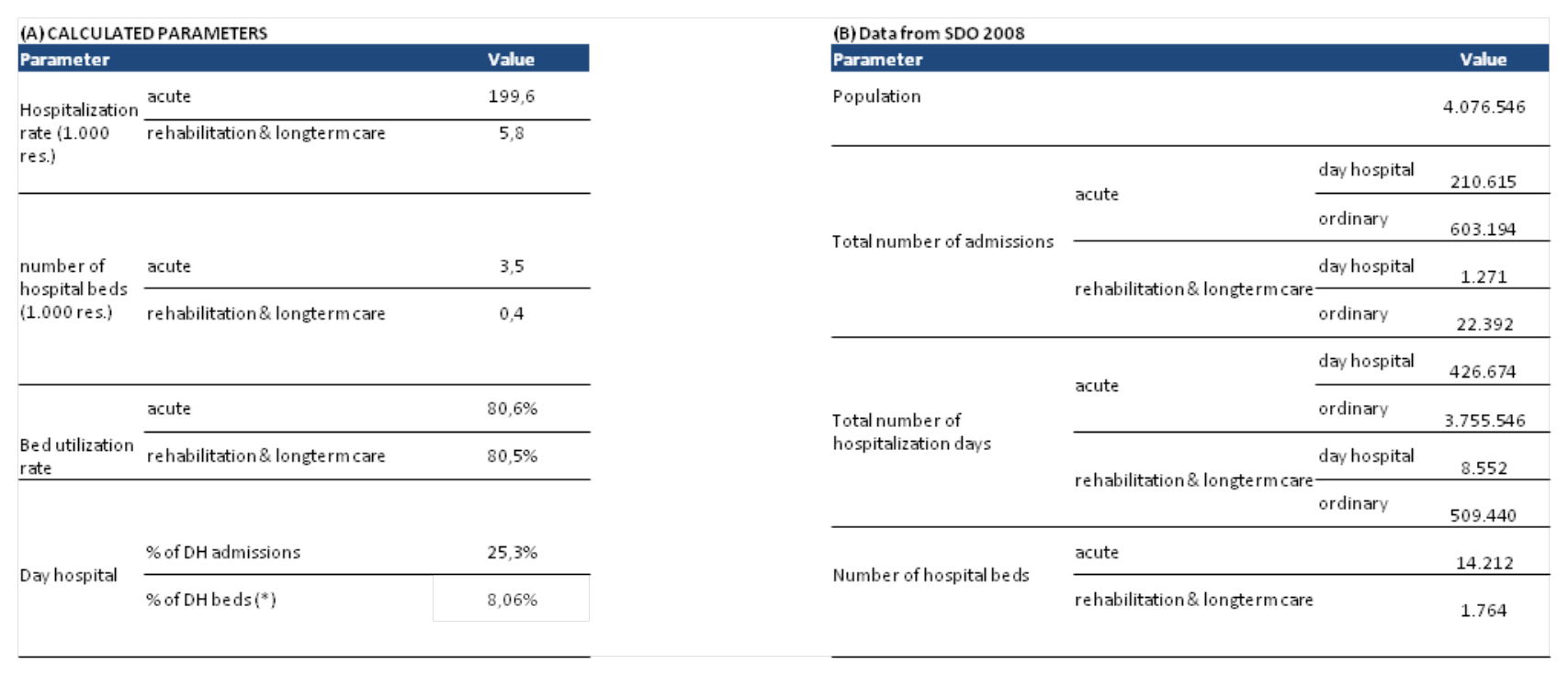}
\caption{Calculation of some parameters (table A) based on data reported in SDO 2008 (table B). \textit{Dati SDO 2008, Ministero della Salute}.}
\label{fig: Osp_12}
\end{figure}

Given that we have to make sure that the re-engineered hospital network meets the legal constraints of section \ref{sec: constraints}, let's start by having a look at how far those parameters are from legal constraints.
Fig. \ref{fig: Osp_11} and \ref{fig: Osp_12} show some data on 2008 hospital admissions taken from the public database of the Italian Ministry of Health. 

\noindent The average \textit{observed} bed utilization rate ($\beta = 80\%$) reported in fig. \ref{fig: Osp_12} has been calculated based on the \textit{observed} total number of hospitalization days ($D$) and the \textit{observed} density of hospital beds ($n$) within the Apulian population ($P$): 
\begin{equation} \label{eq: Osp_2}
\beta_{obs} = \frac{D}{365 \cdot n_{obs} \cdot P}
\end{equation}

\noindent The \textit{observed} total hospitalization rate, $\alpha = 206$ \footnote{Hospitalization rates are reported as number of hospitalized patients every 1.000 residents.}, has been calculated as the \textit{observed} total number of hospital admissions ($R$) over the resident Apulian population ($P$):
\begin{equation} \label{eq: Osp_3}
\alpha_{obs} = \frac{R}{P}
\end{equation}
While the \textit{observed} utilization rate falls within the legal constraints ($75\% < \beta_{obs} < 100\%$), the hospitalization rate doesn't ($\alpha_{obs} > 180$). So apparently we would argue that current Apulian hospital network is affected by a high demand compared to standard levels, which then could at least partially explain the regional healthcare deficit.

However this is only partially true. Let's focus on the last row of fig. \ref{fig: Osp_16}. It shows the calculated system parameters (utilization rate, hospitalization rate, change in hospital beds, etc.) based on current system acute and rehabilitation bed density \footnote{i.e. those numbers reported in fig. \ref{fig: Osp_12}.}. Based on the equilibrium hypothesis of sec. \ref{sec: fundamentals}, if we enforce the current hospitalization rates $\alpha$ and solve for the equivalent bed utilization rate $\beta$, the calculated bed utilization rate is 97,7\%, far above the observed utilization rate of current Apulian hospital network (81\%) reported in fig. \ref{fig: Osp_12}. What does it mean? It means that in an equilibrium scenario, the optimal utilization rate of current hospital network would be much higher than the actual one. It means that there are too few beds to cover the high demand of hospital services such that the only way to work efficiently is to have an extraordinary high utilization rate. \\

\begin{figure}[!htbp]
\centering
\includegraphics[scale=0.77]{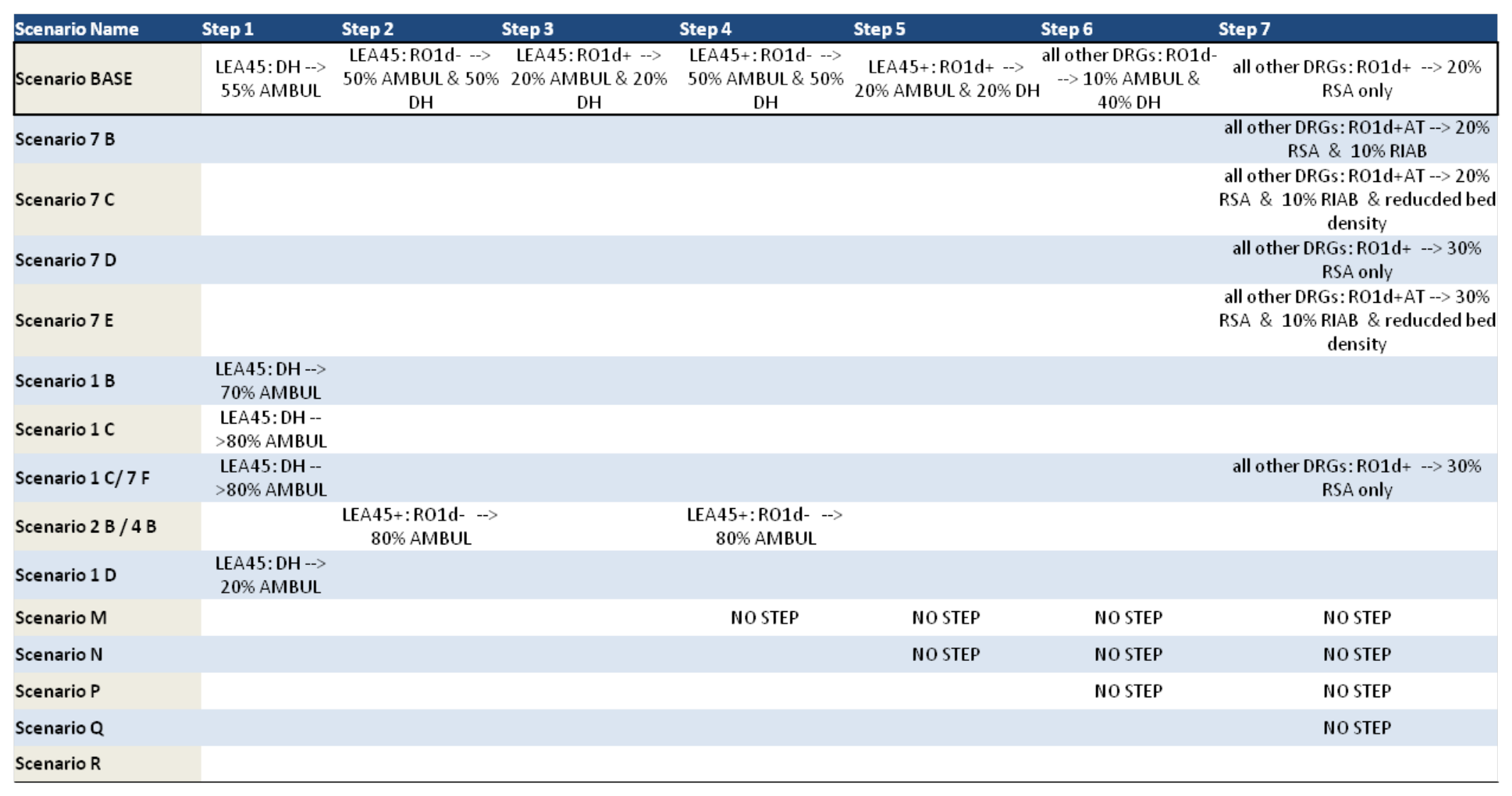}
\caption{The 7 steps of the proposed Base restructuring Scenario are shown in the first row. Starting from the Base Scenario (which is called Scenario R), additional scenarios are generated. Each row describes the differences between the relevant scenario and the base one. When no comments are present it is meant that the relevant step of the base scenario has been applied. E.g. Scenario 2B/4B is made of the following steps: Step 1, modified Step 2, Step 3, modified Step 4, Step 5, Step 6 and Step 7.}
\label{fig: Osp_14}
\end{figure}

After modeling the demand of hospital services according to the historical Apulian demand as reported in SDO 2008 (see sec. \ref{sec: hospitneeds}) and the balanced supply according to sec. \ref{sec: fundamentals}, we have simulated a set of scenarios. The simulated scenarios are reported in fig. \ref{fig: Osp_14}.

A Base Scenario is calculated. It is made of 7 steps. Each step represents a different allocation of hospitalization demand. Fourteen scenarios are then generated by modulating some of the 7 steps of the base scenario.
For example, by reference to fig. \ref{fig: Osp_14} we see that Scenario M is made of three steps only: Step 1, Step 2, Step 3. In terms of hospital admissions each one of these steps has the same allocation of the equivalent steps in the Base Scenario. Let's consider Scenario 1C. We see from the table that its Step 1 has been modulated from the equivalent Step 1 of the Base Sceanario: indeed \vv LEA45: DH $\rightarrow$ 80\% AMBUL" means that 80\% of the day hospital admissions falling within the group LEA45 (see next section) have been reallocated to ambulatory services (as a reference, the equivalent Step 1 of the Base Scenario, reported a 55\% reallocation to ambulatory services).

\subsection{The base scenario}
\begin{figure}[!htbp]
\centering
\includegraphics[scale=1.0]{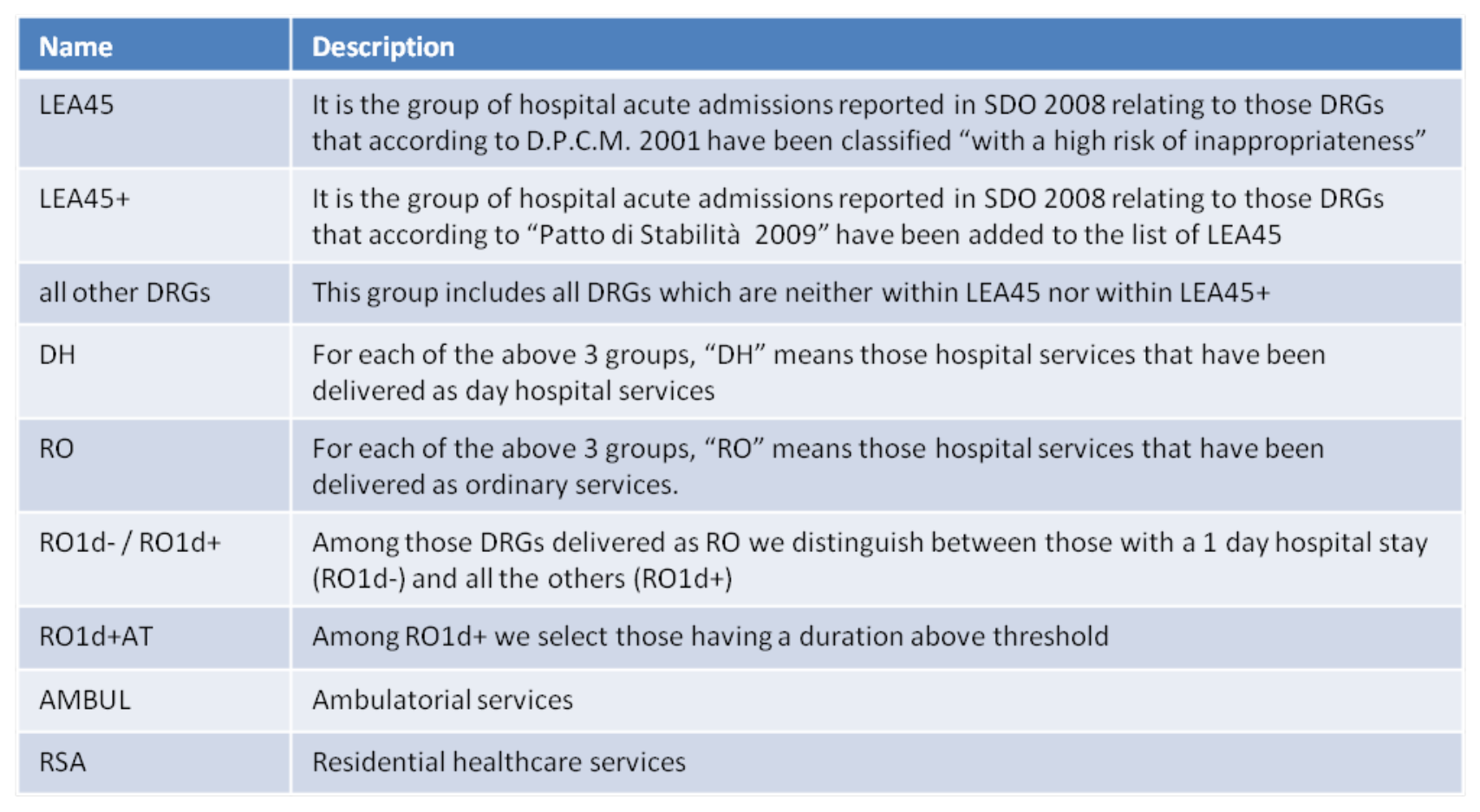}
\caption{Summary of the main short names used in this section.}
\label{fig: Osp_18}
\end{figure}

Let's have a look at how each one of the 7 steps of the base scenario has been modeled. We have grouped the 7 steps in three sets depending on the group of DRGs that have been reallocated.
A comprehensive explanation of the various short names used within this sections is reported in fig. \ref{fig: Osp_18}.

\subsubsection{LEA45 DRGs}
The first two steps reallocate LEA45 DRGs among day hospital and ambulatory services. As discussed in a previous section, LEA45 DRGs are the ones that since 2001 have been classified to be \vv with a high risk of inappropriateness". Under normal conditions they should be delivered as DH or AMBUL services. Recurring to ordinary hospitalization for those DRGs should be avoided and limited to exceptional clinical circumstances.
Given all the above and based on what has been already done (and has been approved by the Government) in other Regions having the same deficit issue of Puglia, we proposed to reallocate those DRGs according to the following:\\

\noindent STEP 1: LEA45 DH  $\rightarrow$ 45\% \textit{DH   and   } 55\% AMBUL \nonumber \\
STEP 2: LEA45 RO1d-  $\rightarrow$ 50\% \textit{DH   and   } 50\% \textit{AMBUL }  \\
STEP 3: LEA45 RO1d+  $\rightarrow$ 20\% \textit{DH   and   } 20\% \textit{AMBUL  and  } 60\% \textit{RO1d+} \\

\noindent Basically, all those admissions (LEA45 DH and LEA45 RO1d-) that should more appropriately be delivered as DH or AMBUL are redistributed among these two categories. A relevant portion (60\%) of RO1d+ admissions is left as RO1d+ : while for 1 day admissions (RO1d-) the DH/ambulatory delivery is by far the most appropriate solution, for long lasting ordinary admissions there could be the chance that clinical conditions make the hospitalization more suitable (e.g. in case of an emergency admission). Another reason for leaving such a relevant portion of RO1d+ as ordinary admission is that even in the best case scenario where none of those admissions has clinical conditions that would make an RO preferable, it is very unlikely the Apulian Region could restructure in one or two years its net of ambulatiory services to divert the full demand coming from these LEA45. It suffice to say that the number of LEA45 RO admissions in 2008 have been 103.155, i.e. a 20\% of the total number of RO acute admissions.
One last consideration is that the proportion of inappropriate RO1d+ DRGs calculated based on the APPRO methodology \cite{APPRO} is 30\%. Based on all these points we chose to take a conservative view and transfer only a 40\% of those admissions.

\subsubsection{LEA45+ DRGs}
Those admissions registered in 2008 as LEA45+ DRGs, based on \vv Patto di Stabilit\`a 2009" between the Italian State and its Regions should be classified as \vv with a high risk of inappropriateness". We aimed at treating the 2008 DH admissions of this group exactly as the ones of LEA45. Unfortunately, no information was available in the report  SDO 2008 on day hospitals admissions relating to those DRGs.
Based on the same conservative assumptions that we adopted for LEA45, we reallocated LEA45+ admissions according to the following:\\

\noindent STEP 4: LEA45+ RO1d-  $\rightarrow$ 50\% \textit{DH   and   } 50\% \textit{AMBUL }  \\
STEP 5: LEA45+ RO1d+  $\rightarrow$ 20\% \textit{DH   and   } 20\% \textit{AMBUL  and  } 60\% \textit{RO1d+} \\

\subsubsection{All other DRGs}
 As far as the DH admissions are concerned, no information was available, hence we conservatively assumed they are left as DH (the alternative would have been to divert them to ambulatirial services). As far as the RO1d- group is concerned, a 40\% portion of them has been reallocated to DH services. Based on data reported in \cite{RE_Hampton, RCPhys, RE_Raza, RE_Hanlon, RE_Stewart} we assumed that another 10\% reduction could take place as a consequence of the introduction of appropriate bed management measures such as the Acute Medical Admissions Unit (AMAU).

\begin{figure}[!htbp]
\centering
\includegraphics[scale=0.7]{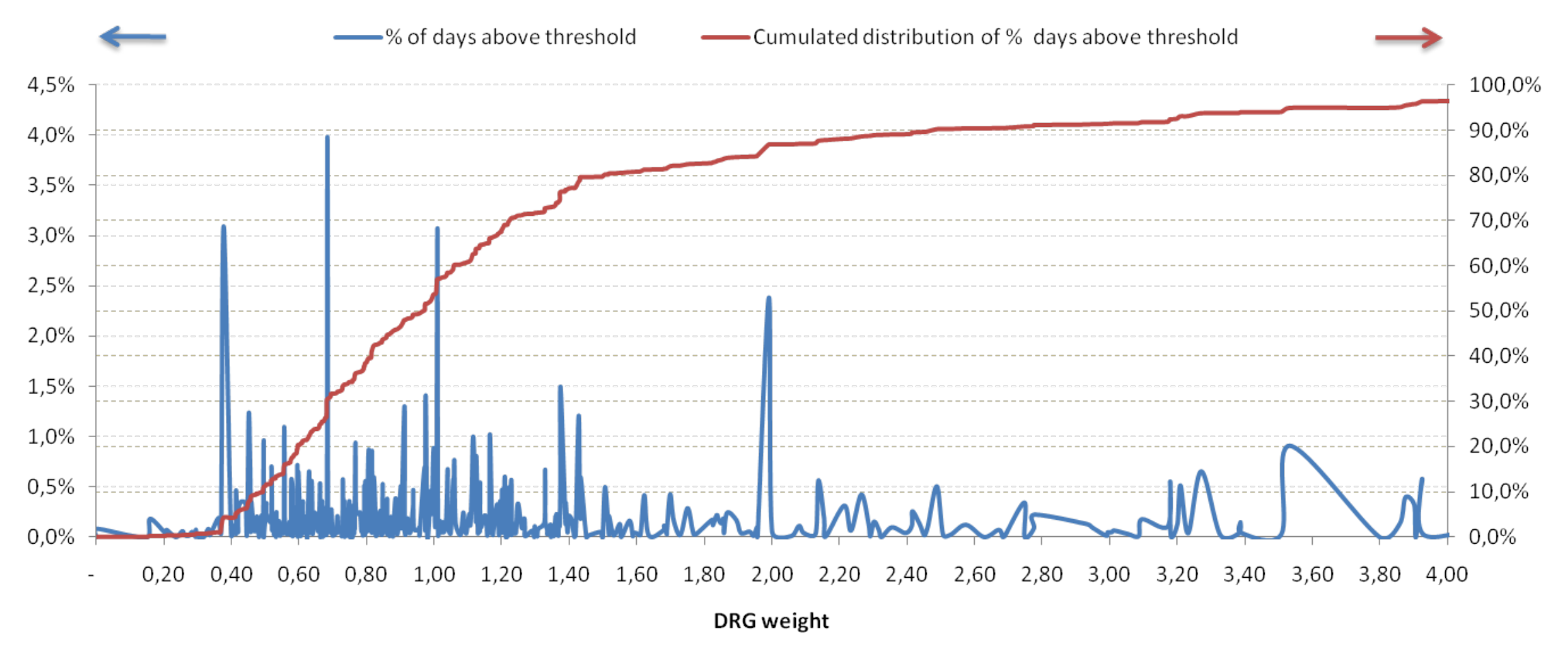}
\caption{Distribution of days above threshold among all DRGs delivered as ordinary acute admissions. For each DRG we reported its DRG-weight and the associated percentage of days above threshold. Percentages are calculated by dividing the number days above threshold by the total number of days above threshold. The chart demonstrates that the most part of days delivered above threshold are associated to DRGs with a relatively low weight. \textit{Source: SDO 2008 report, Ministero della Salute}.}
\label{fig: Osp_15}
\end{figure}

All other DRGs are not classified with a high risk of inappropriateness. Hence no rellocation was done on RO1d+ among DH and AMBUL. However, a 20\% trasfer of RO1d+AT to alternative hospitalization services, such as RSA \footnote{Residential and semiresidential services.} and home care, was suggested. Indeed, as fig. \ref{fig: Osp_15} suggests, a 40\% of the total number of days above threshold are associated to DRGs with a low DRG weight. The DRG weight is a measure of the intensiveness of the healthcare service that has been delivered in association to that DRG. 

Another relevant data comes from the percentage of above threshold SDOs \cite{SDOdef} associated to elderly people (i.e. above 65 years): based on 2008 data, 3,5\% of acute ordinary admissions are above threshold. When one takes into account the realized hospitalization rates for elderly people \footnote{65-74 cohort: 325 (males) and 230 (females). 75+ cohort: 475 (males) and 335 (females). Based on SDO 2008.} one sees that these amount to roughly a 20-30\%  of the total number of acute ordinary admissions above threshold.
\noindent The chosen reallocation for all other DRGs follows:\\

\noindent STEP 6: all other DRGs RO1d-  $\rightarrow$ 40\% \textit{DH   and   } 10\% \textit{AMAU }  \\
STEP 7: All other DRGs RO1d+AT  $\rightarrow$ 20\% \textit{RSA }\\

As it is shown in fig. \ref{fig: Osp_14} alternative solutions to full RSA reallocation of those RO1d+AT admissions have been modeled in some of the other scenarios. The idea is to allocate a percentage of them to rehabilitation and longterm care services. Being rehabilitation services more costly than residential ones, of course this solution comes with higher costs. Moreover, the feasibility of this alternative will depend on the actual needs of Apulian patients (rehabilitation services are not equivalent to RSA services from a healthcare point of view).

\section{Results}
\label{sec:Results}
\begin{figure}[!htbp]
\centering
\includegraphics[scale=0.75]{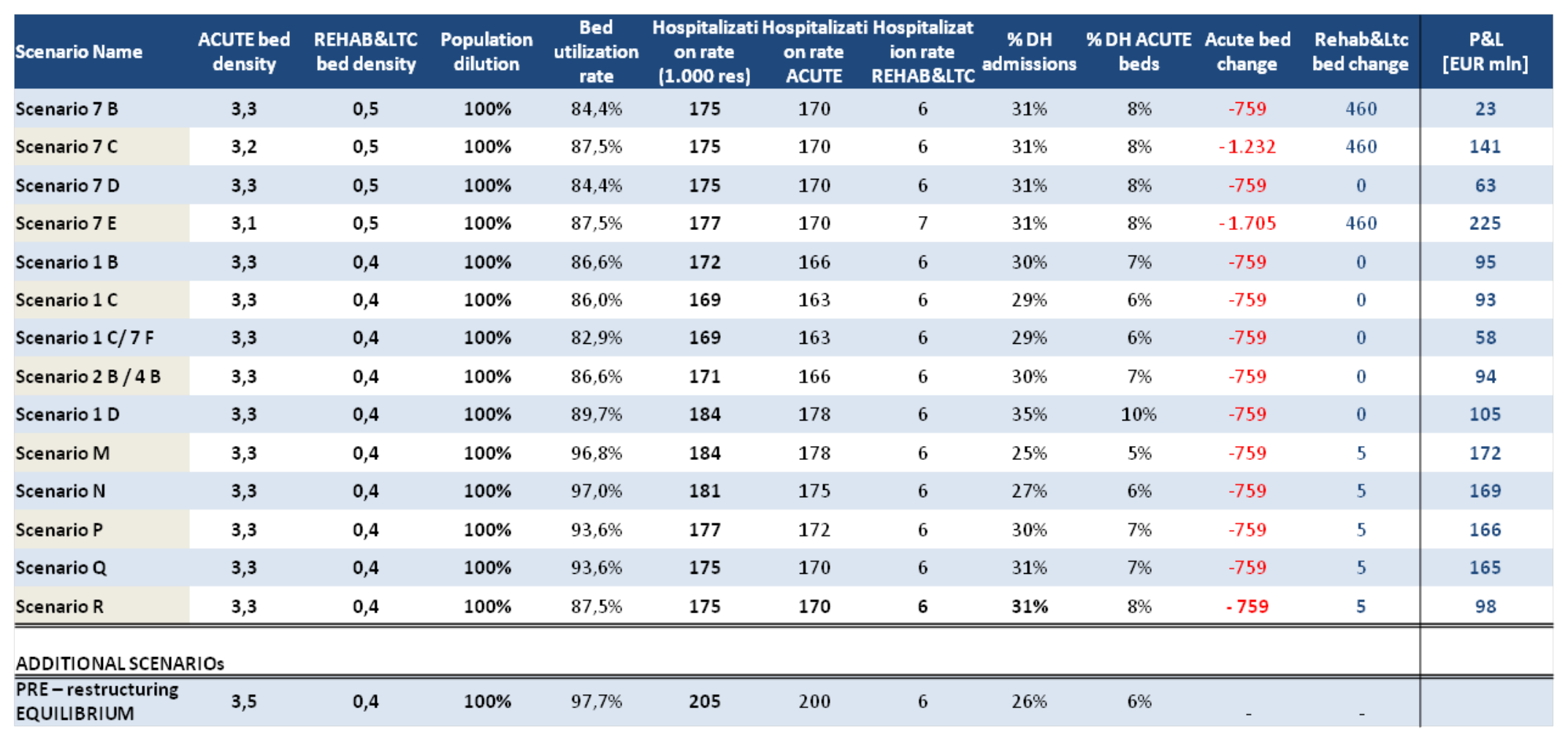}
\caption{Different scenarios have been simulated. The table shows how the various system parameters changes among the different scenarios. The last column on the right reports the expected financial P\&L associated to bed cuts. The last additional scenario is not proposed as alternative to hospital network restructuring but is used to assess current system unbalances (see text).}
\label{fig: Osp_16}
\end{figure}

%
\noindent 
The fourteen scenario generate the different results shown in fig. \ref{fig: Osp_16}. By looking at that table we see, for example, that Scenario 1C assumes a density of acute beds of 3,3 (the legal constraint), a density of rehabilitation and longterm care beds of 0,4 (which is below the legal constraint), a hospitalization rate of 86\%, and so on. The final column reports the P\&L of each of the 14 scenarios. 

\begin{figure}[!htbp]
\centering
\includegraphics[scale=0.8]{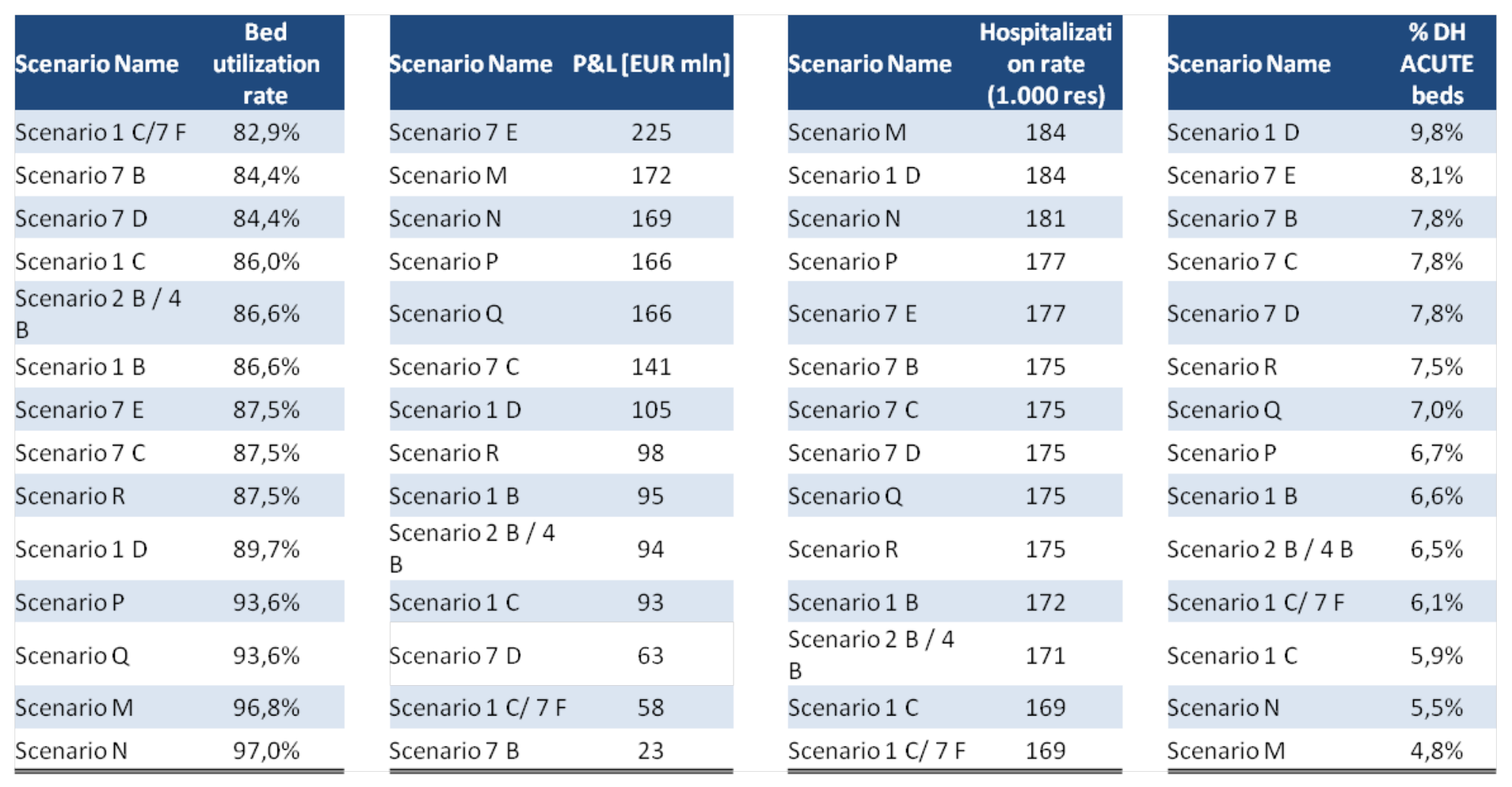}
\caption{Ordering of the 14 scenarios by four parameters: hospitalization rate, bed utilization rate, P\&L and \% of acute beds.}
\label{fig: Osp_19}
\end{figure}

What scenario is the best one?\\
\noindent The answer is \vv it depends". Indeed, it depends on the non-legal constraints that we have. 
For example, we could be driven by financial constraints (which is the case for Regione Puglia) and by having a look at fig. \ref{fig: Osp_19} we see that Scenario N and Scenario 7E are the best ones in terms of generated value (P\&L). 

\noindent However, by choosing Scenario N, we are actually saying that we are able to deliver a bed utilization rate of 97\%, well above the current level of 81\%!\\
Alternatively, by choosing Scenario E7 looks even brighter in terms of saved money (EUR 225 mln). However, should we choose to implement a reorganization of the hospital network based on this scenario, the first question that we have to ask is: could we bear a reduction in the total number of beds to 3,1 every 1.000 res. (i.e. below the already tight legal constraint of 3,3)?\\
We are not saying that it cannot be done, but that an additional careful analysis of the overall system is actually needed. A limited physical infrastructure or a lack of organizational management pose several limits on the capability of the system to be re-engineered according to those solutions having a high bed utilization rate. A not homogeneous distribution of hospitals could be a major factor hindering the enormous cut of 1.707 beds foreseen by Scenario E7. 

\begin{figure}[!htbp]
\centering
\includegraphics[scale=0.9]{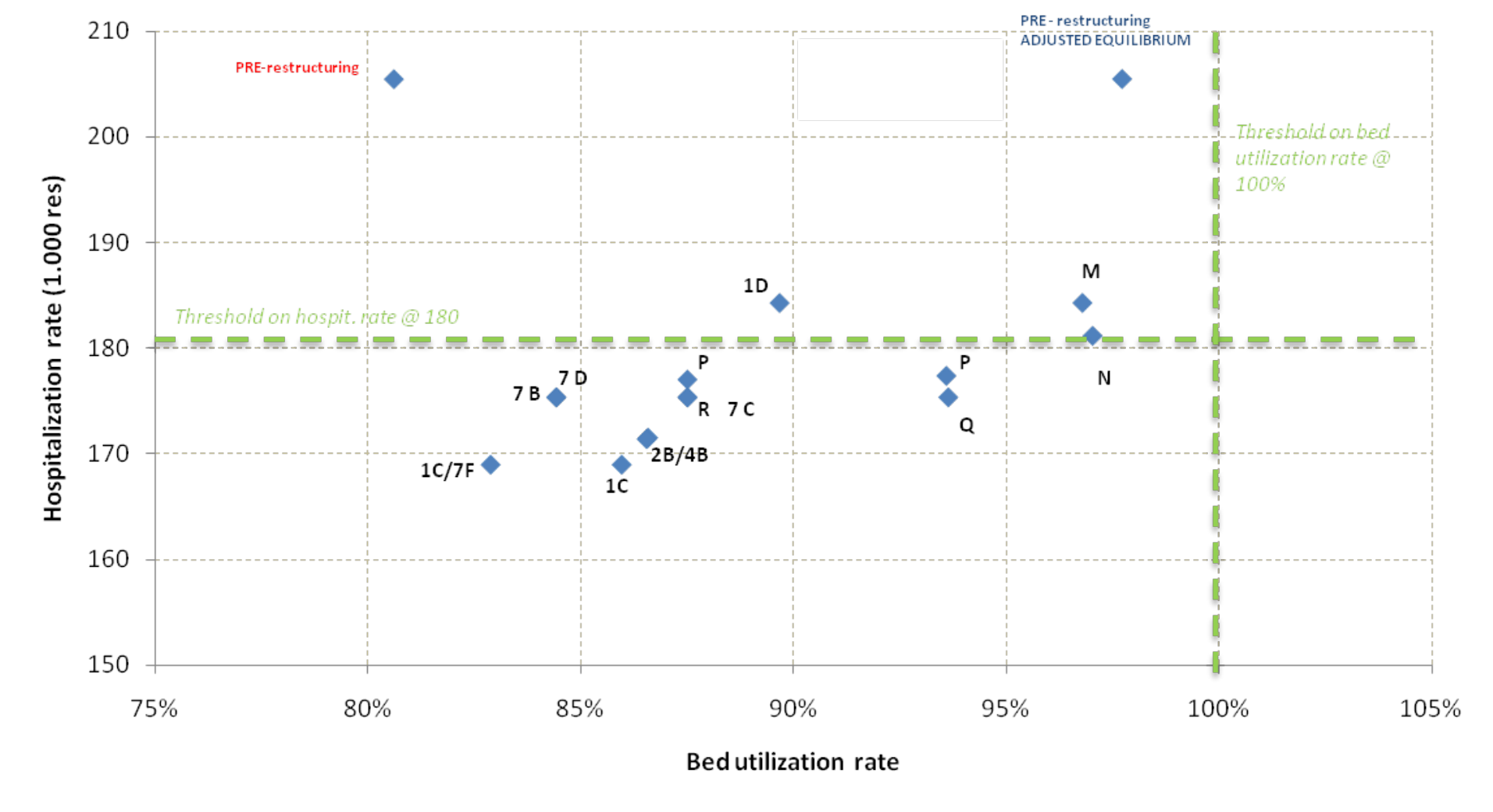}
\caption{The plot shows the system parameters (hospitalization rate and bed utilization rate) corresponding to the various step described in fig. \ref{fig: Osp_14}.}
\label{fig: Osp_13}
\end{figure}

A graphical representation of the evolution of the system throughout the various scenarios is given in fig. \ref{fig: Osp_13}.
\\
We think that a good starting point for an advanced planning would be Scenario 1D and Scenario R, i.e. according to the roman proverb \vv in medio stat virtus", we chose two \vv average" scenarios. \\

Both scenarios require a cut of 759 hospital beds. Having to choose between the twos, we think that the easier to implement would be the second one: indeed, maybe it is easier to increase the utilization rate by two additional points (from 87,5\% to 89,7\%) than to reduce the hospitalization demand (from 184 to 175).\footnote{One last note on fig.\ref{fig: Osp_16}. The reader will see that most of the scenarios seem to not qualify for the constraint on the percentage of day hospital beds. As the following two arguments will show, this is not the case. The first argument is that the 10\% constraint applies to the overall DH beds, while the reported number refers to acute beds only. The second argument is that the current public sector of the hospital network (which is the only one where we have more reliable data) is working with a percentage of DH beds below 10\% (around 8\%). To this extent, the proposed solutions are in line or improve the current system. Unfortunately we cannot calculate the ratio for the overall system given that we have not been able to find the actual split between DH and RO private beds (neither SDO 2008, nor the Ministry of Health website provided any useful information) and without this number there is a lot of uncertainty in the expected fraction of DH beds as well. We will be back on this in the next section.}

\begin{figure}[!htbp]
\centering
\includegraphics[scale=0.7]{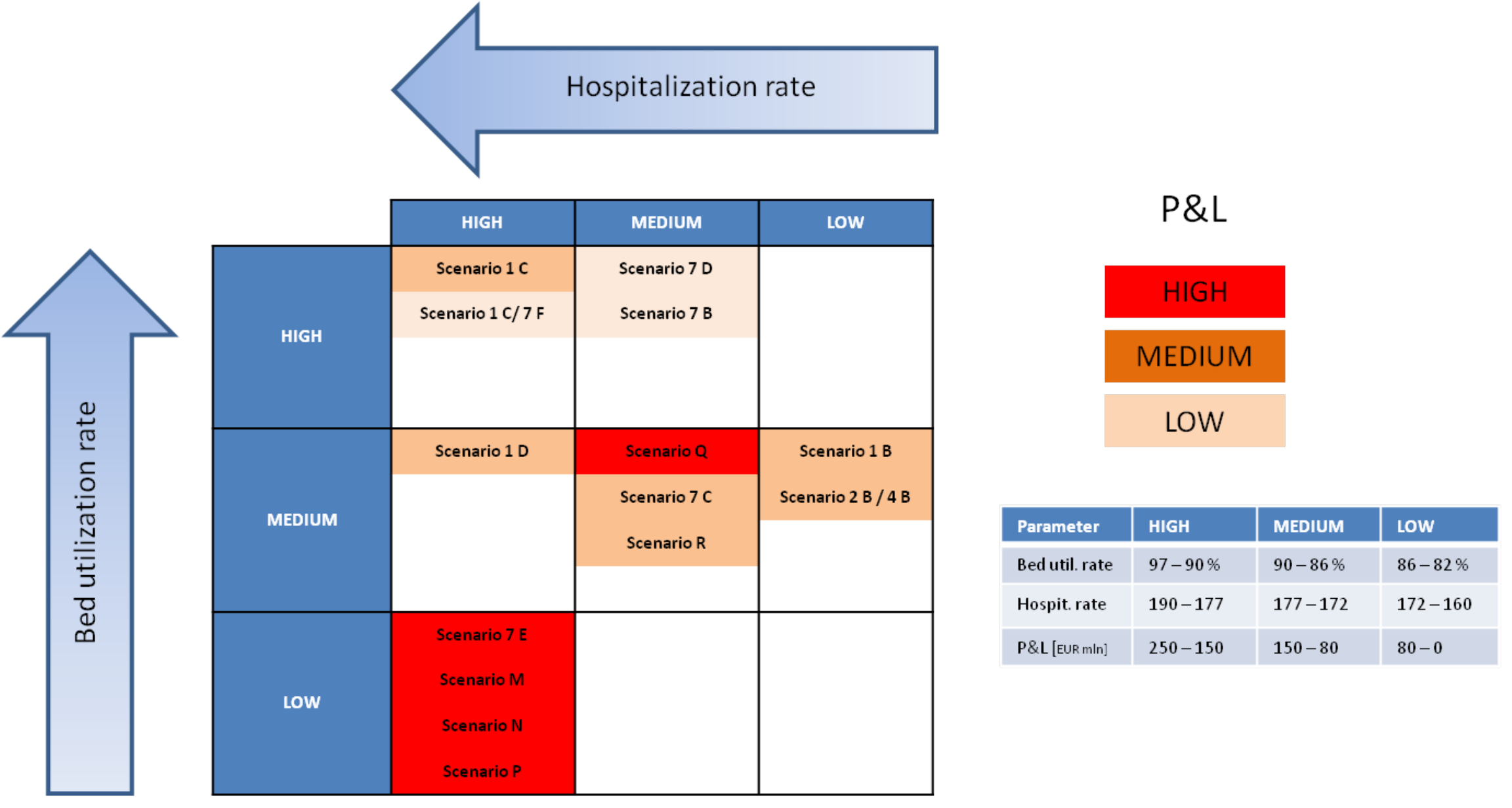}
\caption{Grouping of the simulted 14 scenarios according to three parameters which are very relevant for the planning of the restructured hospital network. The table on the right shows the ranges used for each parameter.}
\label{fig: Osp_17}
\end{figure}

To conclude this section, in fig. \ref{fig: Osp_17} we report a useful grouping of the different scenarios.
As already stressed, unfortunately we are not in the position to advise on a particular decision: the final aim of simulating different scenarios has been to provide a sensitivity of the hospital network to the various different design-parameters and to allow the recipient of this work (ideally whoever will be in charge for the restructuring of a hospital network) to assess which one was better suited to the needs of the Apulian population.
The actual selection of the \vv best" scenario is unfortunately beyond the scope of this work.


\section{Limits and perspectives} 
\label{sec: limitsper}
%
\noindent 
How good is our planning? How much confident are we about our numbers? What improvements could be done?
These type of questions will be answered in this section. The reader will see that there are two real limits to what can be done: data and time. 

Data are not that easy to find as one could expect. Moreover sometimes it happens that even if you find the data, unfortunately they are not consistent. The more informations you have, the more features of the problem you are able to assess. Your planning will get even closer to the best solution available. Indeed, some interesting topics have not been covered just because we were missing enough details on data.

As far as time is concerned, clearly one has to consider that in practical terms one has not an infinite time to find a solution. The right solution is the best one that you can find within the limited time available. For example, it would be interesting to see what could be the impact on the proposed restructuring of future possible epidemiologic scenarios. However much more time would be needed to find a proper solution to that. 

At the end of the day, we preferred to deal with the best set of data we could publicly access through what is considered the official source of informations related to the Italian NHS. 
We are sure that allowing for more time and for additional informations, the results that we have shown could be improved. \\

Not all the answers provided in this last section will be exhaustive, some of them want to be more properly considered as proposals for future developments of the work presented in this paper.

\subsubsection*{LEA and LEA45+ DRGs}
As previousely explained no information was available in the report  SDO 2008 on day hospitals admissions relating to LEA45+ DRGs and as a consequence we were not able to \vv extract value" by reallocating them among day hospital and ambulatory services. Given the relevant portion of LEA45+ DRGs on the overall number of SDOs, we think that it would be worth to investingate their potential impact as soon as additional data from Ministero della Salute become available. 

Moreover the 60/40 split among RO1d+ DRGs of both LEA45 and LEA45+ groups could seem a bit conservative. Previous reports relating to the restructuring of the hospitalization net of other Italian Regions seem to point in the same direction: maybe we have been too much conservative. However a better understanding of the inappropriateness of those admissions would need additional informations. Given that no additional information is at our disposal, we preferred to be on the safe side. The relevant proportion of both LEA45 and LEA45+ (40\% in total) to the overall number of DRGs justifies additional future work on the subject.  

\subsubsection*{Private hospitals}
There are two informations that are missing in relation to private hospitals.\\
The first one is the number of day hospital beds among private hospitals. Given that the same data are reported for public hospitals and we know that DH beds among public hospitals for \text{acute} cares amount to 1.008, in order to calculate the number of \textit{acute} DH beds among private hospitals it would suffice to know the total number of DH beds of the RHS. However this number is not reported \footnote{And it cannot be derived based on the other data provided.} as well. \\

Unfortunately we have to resign ourselves that no information at all is available in relation to the total number of DH beds. 
While it obviously makes no difference as far as the absolute new allocation of beds is concerned, the lack of information on private splitting between DH and RO affects any differential effect between pre-restructuring and post-restructuring: the calculated changes in hospital beds and the net financial impact of the restructuring are two of the main ones. \\

Given that one could be interested in assessing which specialty needs bed-cuts and which other needs bed-additions and given that any plan of this kind is almost useless without having an estimate of its financial impact, we had to make an assumption.
Indeed we estimated the number of acute private DH beds by first estimating the total \footnote{\vv total" meaning both public and private beds.} number of DH beds. The last one has been calculated based on reported number of acute DH accesses (again, both public and private accesses):
\begin{equation} \label{eq: Osp_1}
\textit{number of DH beds} = \frac{ \textit{total number of DH accesses} }{A \cdot f \cdot 250 \cdot \beta}
\end{equation}
where $A=2$ and $f=100\%$ \footnote{Some documents, as the one relating to the restructuring of the hospital network of Regione Lazio, based on an a statistical analysis on the usage of day hospital beds, report that a correction factor of $f = 75\%$ should be introduced when calculating day hospital beds. We used this adjustment when estimating current day hospital beds.} are already known parameters and $\beta$ is the utilization rate. A 80\% utilization rate has been asssumed consistently with data reported in fig. \ref{fig: Osp_12}. The total number of acute DH beds was calculated to be 1.080. Based on this and on the number of acute public DH beds (985) we can calculate the estimated number of private acute DH beds (23). This in turn allows us to estimate the number of ordinary beds for private hospitals. By adding it to the number of public ordinary beds we get an estimate of the total number of ordinary beds. We are then able to calculate the differential in ordinary hb between post- and pre-restructuring.
This data is then used to calculate the financial impact.
\begin{figure}[!htbp]
\centering
\includegraphics[scale=0.95]{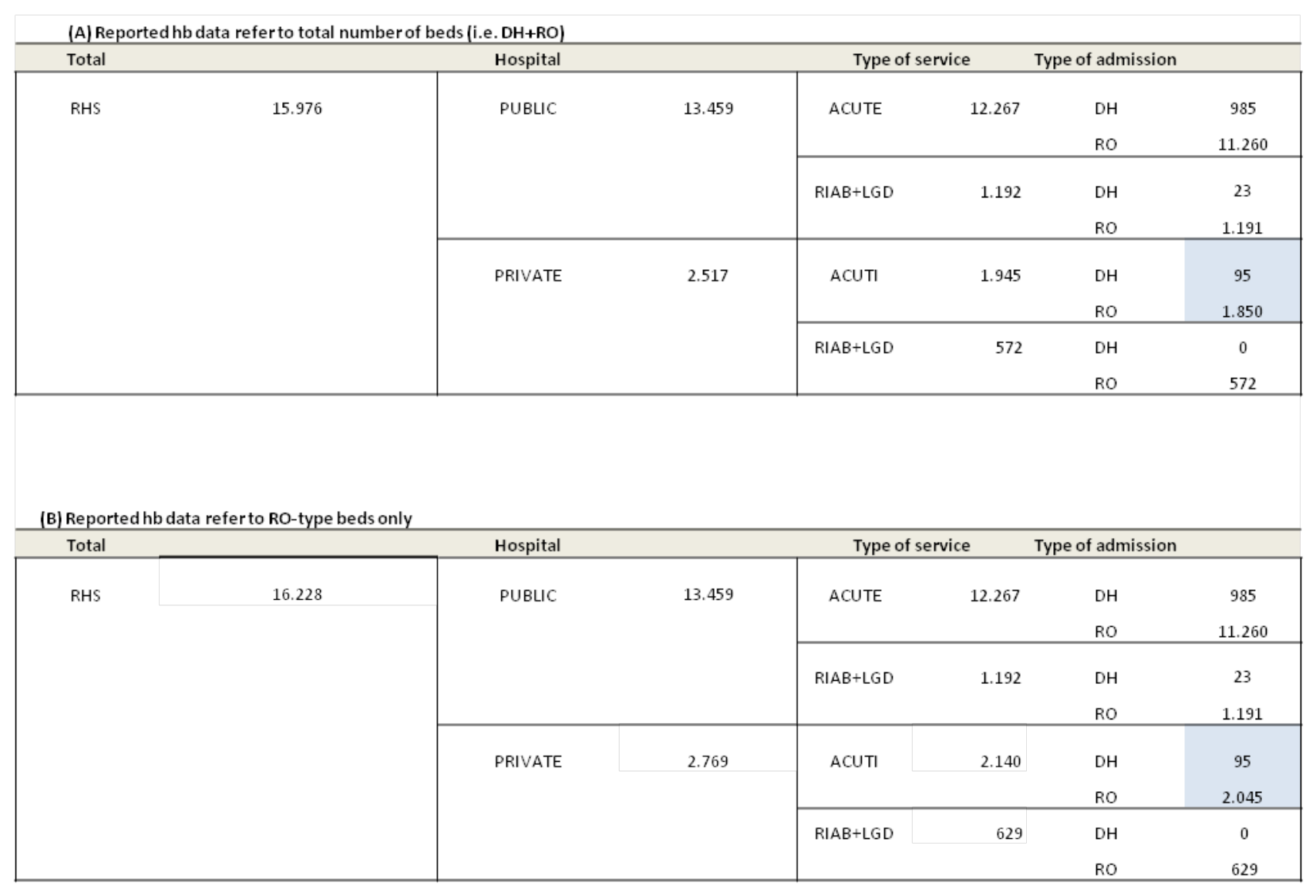}
\caption{Distribution of beds among public and private hospitals. The cells in light blu are estimates based on SDO 2008 number of discharges and lenght of hospital stays. The two alternatives (A) and (B) correspond to numbers shown in fig.\ref{fig: Osp_6}.}
\label{fig: Osp_10}
\end{figure}

The second information which is missing has already been introduced in section \ref{sec: constraints}.  As it is shown in fig. \ref{fig: Osp_6}, the estimate of total number of acute beds (see eq. \ref{eq: Osp_1}) allows us to have the allocation of beds among public/private and DH/RO based on the two different assumptions reported in fig. \ref{fig: Osp_6}. \\

\subsubsection*{Bed allocation among specialties}
\begin{figure}[!htbp]
\centering
\includegraphics[scale=0.95]{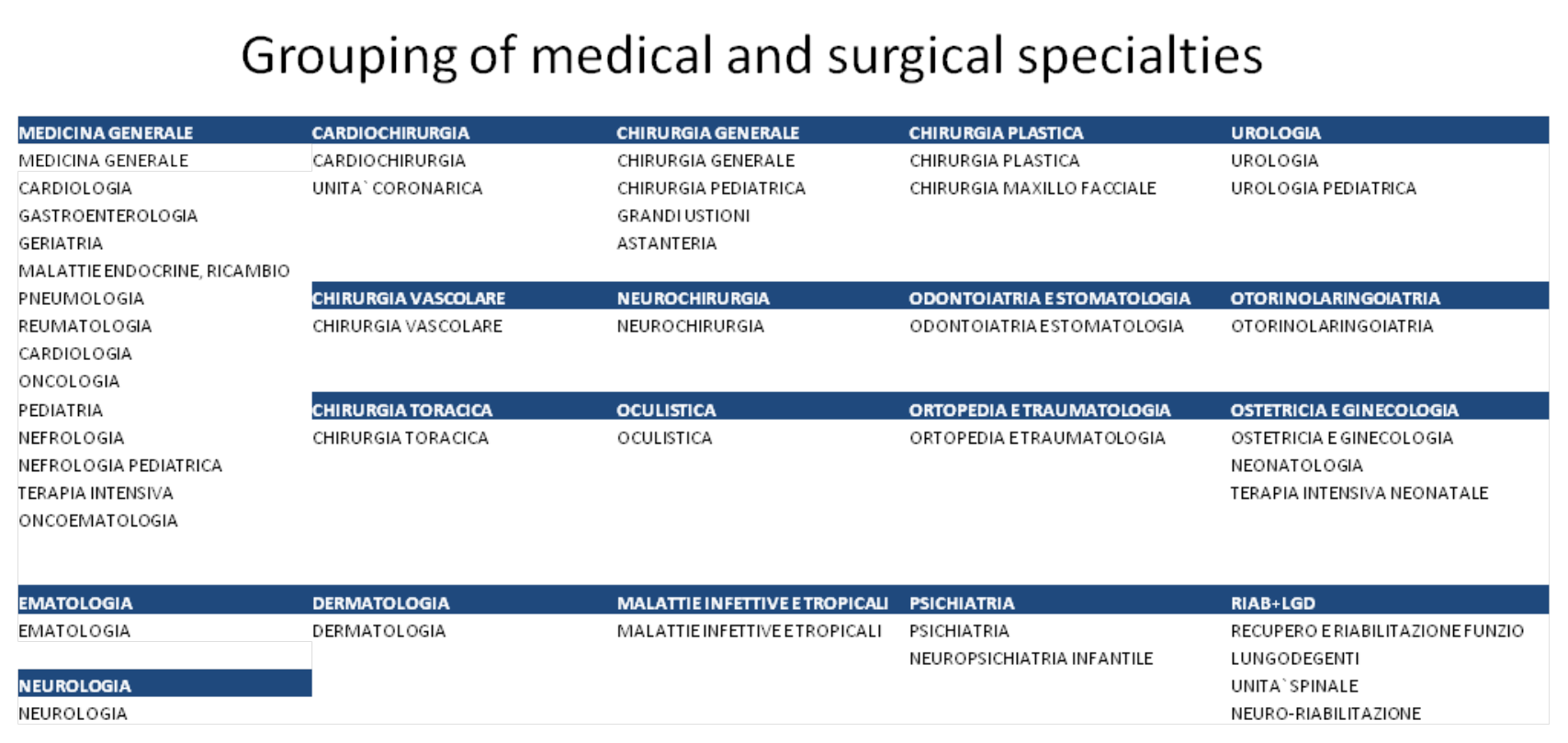}
\caption{Grouping of those medical and surgical specialties that have been reported in 2006 report of Ministero della Salute on \textit{public} hospitals. Each group contains a set of specialties sharing a high level of affinity in terms of DRG delivered. For example: DRGs coming from a Geriatric departments are more likely to be delivered by a General Medicine department than from a Vascular Surgery one.}
\label{fig: Osp_21}
\end{figure}

An interesting and necessary next step of our plan of the re-engineering of the hospital network would be to determine the new allocation of the hospital beds among clinical specialties. Unforunately, due to lack of data relating to the split of current beds among clinical specialties of \textit{private} hospitals we are not able to perform such planning.
Nevertheless, we decided to see if we could have been able to extract any valuable information based on the only data at our disposal. 
\begin{figure}[!htbp]
\centering
\includegraphics[scale=0.95]{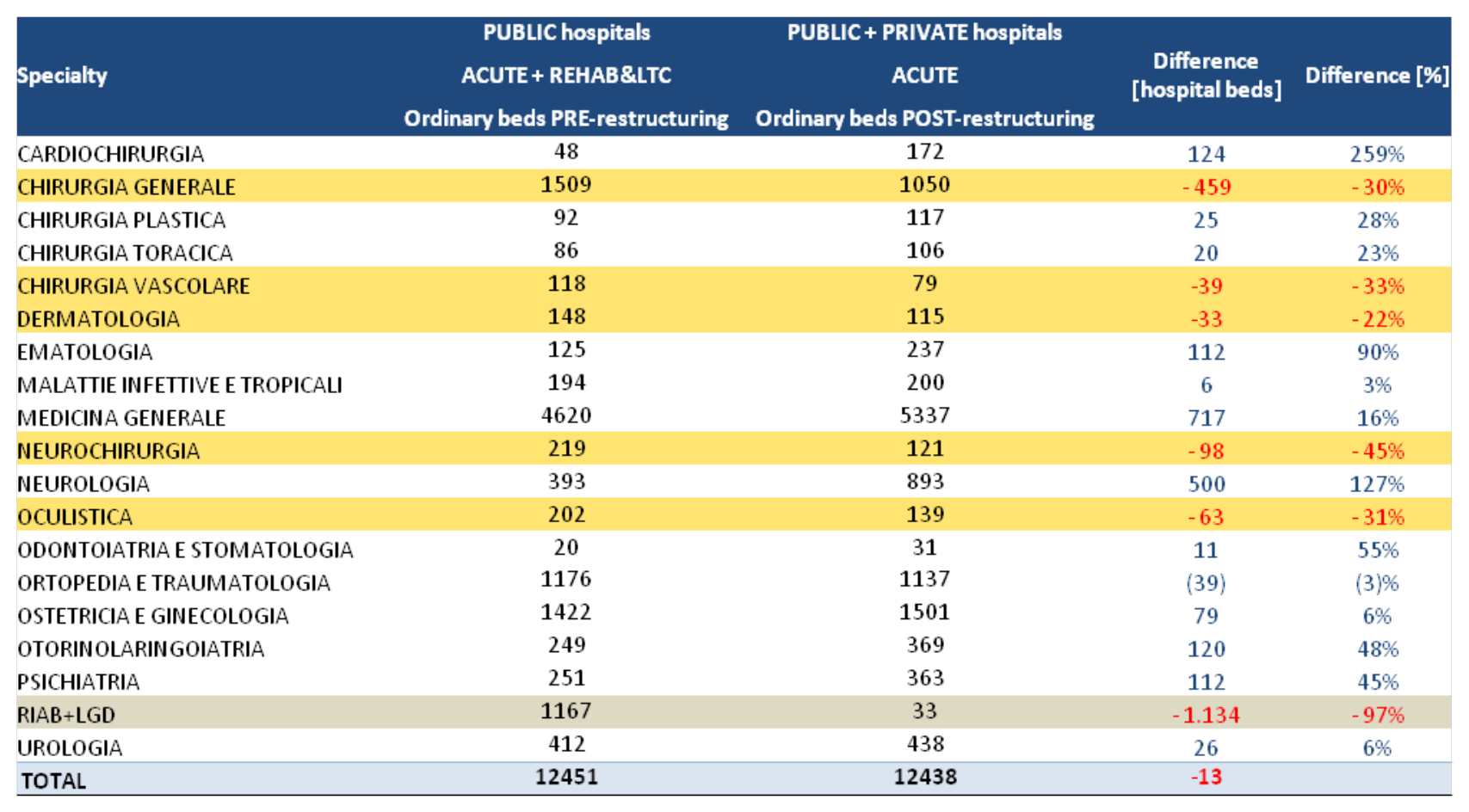}
\caption{Comparison of the bed allocation among clinical specialties. Both columns refer to ordinary beds. Please note that while the first column refers to the total number of beds currently available among \textit{public} hospitals \bb{only}, the second column refers to the \textit{total} (public and private) beds that are planned as a consequence of our restructuring. Moreover, while the first column refers to both \textit{acute and rehabilitation and longterm care} beds, the second refers to \textit{acute} beds \bb{only}. The third column reports the net percentage change between the two columns. \textit{Source: current beds have been taken from SDO 2006, available at Ministero della Salute public website}.}
\label{fig: Osp_20}
\end{figure}
The analysis was made along the following lines:
\begin{itemize}
\item  we have selected all those specialties which were shown in a 2006 report on \textit{public} hospitals publicly available at Ministry of Health website \footnote{Unfortunately the 2006 report was the most updated one.}. Within that report, for each public hospital it was reported the number of working beds available for each specialty unit available at that hospital;
\item  we grouped those specialties in major sets following a criterion of DRG affinity: for example Endocrinology and General Medicine are closer than Endocrinology and Chest Surgery, i.e. it would be much more likely to find the same DRGs delivered by an Endocrinology and a General Medicine departments than it would be between Endocrinology and Chest Surgery. Fig. \ref{fig: Osp_21} shows the grouping among specialties;
\item  we attributed each of the DRGs available for Puglia in the Ministry of Health SDO 2008 report \footnote{The same we used to assess Apulian demand for acute ordinary hospital cares.} to a particular group of specialties based on a qualitative criterion of affinity: for example a DRG related to some kind of respiratory disease (e.g. BPCO) was associated to the medical specialty of Pneumology, which in turn, based on the specialty grouping falled under \vv General Medicine";
\item  we then calculated the number of beds that would be needed in the new restructured hospital network, specialty by specialty. Formula \ref{eq: fundam} was used with the 87,5\% utilization rate calculated for Scenario R (the Base Scenario) and the associated re-engineered admissions (see row of Scenario R in fig. \ref{fig: Osp_14});
\item  the final step was a comparison, specialty by specialty, of the number of new beds with the number of working beds which were counted within the aforementioned 2006 report on public hospitals.
\end{itemize}

\noindent Results are shown in fig. \ref{fig: Osp_20}. The two columns are not really comparable since they refer to different things: 
\begin{itemize}
\item 	the first one refers to \textit{public} hospitals only: we were able to find bed allocation among specialties for public hospitals only. It shows acute, rehabilitation and longterm care beds alltogether.
\item    the second one refers to \textit{public and private} hospitals and to \textit{acute} admissions only: this happens because the new beds are calculated based on re-engineered SDO 2008 ordinary acute admissions.
\end{itemize}
Even if the two are not immediately comparable, there is something that can be inferred from those data. If we take a look at those specialties with a huge negative percentage difference in number of beds, we see that they all share a common feature: they are all surgical specialties (General Surgery, Vascular Surgery, Neurosurgery and Oculistics). Why is that? A possible explanation is that there are too many medical DRGs coming from surgical departments, which it's a clear sign of \bb{organizational inappropriateness}. This is a valuable information for re-engineering the new hospital network since it means that the next step would be to take a closer look at the what hospitals are affected by such inappropriateness and to plan the right measures to reduce it.
Additionally, we would like to stress that while such valuable information is possible for those specialties with a high negative percentage difference, it is not possible to infer any valuable information for those with a positive percentage change. The reason being that when taking private beds into account (i.e. if we were able to know the current bed allocation of private hospitals) the positive number could reduce its absolute value and eventually turn negative.

\subsubsection*{Planned vs. emergency admissions}
Unfortunately data available on public Italian Ministry of Health database lack some additional information that could turn very useful when planning to restructure a hospital network: the split between \textit{planned} and \textit{emergency} acute admissions. 

It could turn very useful since there is a good portion of \textit{planned} admissions, currently delivered as ordinary admissions, that should more appropriately be delivered as day hospital admissions. The availability of such information for each hospital (both public and private) and for each specialty division (e.g. General Medicine, Dermathology, etc.) would allow one to compare the different regional hospitals. A measure of \bb{appropriateness} could then be established. For example, one could plan that for each DRG the percentage of acute admissions delivered as RO should not be higher than a choosen threshold with a certain tolerance based on statistical distribution of all regional hospitals.\\

Let's see how one could practically do it. As stated so far, the underlying assumption is that a certain portion $P_{i}$ of the planned admissions which are normally managed through ordinary admissions could actually be \bb{more appopriately} managed through day hospital admission.
A performance index can be calculated for each specialty division (e.g. Cardiology, Emathology, etc.) based on the difference between the \textit{expected} and the \textit{observed} number of beds. The \textit{expected} number of hb is calculated as the equivalent number of hb that would be needed if a proportion $P_{i}$ \footnote{$i$ is an index running on the different type of DRGs that are delivered by the specialty division} of total hospitalization days of planned admissions ($D_{i}^{plan}$) would be delivered as DH and the remnant $1-P_{i}$ would continue being delivered as DH.
Emergency admissions (whose total number of hospitalization days is $D^{emg}$) continues to be delivered as RO.
The calculation would be performed as:
\begin{eqnarray}
PI &=& HB_{exp} - HB_{obs} \\
      &=& HB_{exp}^{DH}+ HB_{exp}^{RO}	- HB_{obs}  \nonumber \\
      &=& \frac{D_{DH}}{A \cdot 250 \cdot \beta} + \frac{D_{RO}}{365 \cdot \beta}  -  HB_{obs}\nonumber \\
      &=&  \Big [  ( \sum_{i}  D_{i}^{plan}) \cdot (1-P_{i}) \Big] \cdot  \Big ( \frac{1}{A \cdot 250 \cdot \beta} \Big ) \nonumber \\    
      &+&  \Big [  D^{emg}     +  (\sum_{i}  D_{i}^{plan}) \cdot P_{i} \Big]                  \cdot  \Big ( \frac{1}{365 \cdot \beta} \Big )   - HB_{obs} \nonumber
\end{eqnarray}
where $A = 2$ is the number of same day admissions per bed, 250 and 365 are the conventional number of days adopted for DH (which are working 5 days a week) and RO (which are working 7 days a week).

\subsubsection*{Interregional healthcare mobility}
For 2008 625.048 SDOs have been attributed to acute ordinary admissions of Apulian residents. However, 44.313 among Apulian residents acute ordinary admissions are made outside Puglia (\bb{negative mobility}), while 22.459 hospital acute ordinary admissions refer to people non resident in Puglia \footnote{Mainly coming from the surrounding Regions of Basilicata and Molise.} (\bb{positive mobility}).
The net number of admissions is 603.194, as reported in fig. \ref{fig: Osp_11}.
We want to make three notes on this subject.

\textit{First}, we now understand that the number reported as \vv number of acute ordinary admissions" \footnote{And all the numbers that are reported on SDO 2008 in relation to each DRG.} is incorrectly said to refer to admissions of Apulian residents. Actually it includes non-Apulian residents that have been admitted by Apulian hospitals and excludes the Apulian residents that have been admitted to non-Apulian hospitals.

\textit{Second}, the difference between negative and positive mobility is a relevant number, since it represents a 7\% of the total number of patients having been admitted to Apulian hospitals (i.e. 603.194). This number is not negligible at all, meaning that there are more Apulian residents being cured outside Puglia than viceversa. 

\textit{Third}, while the constraint on the demand for hospital services (hospitalization rate) refers to the total number of admissions registered by Apulian hospitals, the constraint on the offer of hospital services (number of beds every 1.000 residents) is calculated based on the resident population. There is an inconsistency on the standards enforced by the Legislator. If the negative inter-regional mobility becomes a relevant portion of the intra-regional demand, there is a \bb{dilution effect}: the effective population demanding for regional hospital services is less than the one on which hospital beds are \vv offered".

\subsubsection*{Financial estimates}
The numbers reported in the previous sections under the name \vv P\&L" represent the pure profit/loss of bed reallocation. A proper financial assessment should take into account other factors that could be involved when realizing the real re-organization of the hospital network. Just to mention, one relevant factor would be the restructuring of the work-force: clearly, if in the real plan we will decide to close a hospital, the employees (doctors, nurses, administratives, etc) could be reallocated somewhere else within the RHS or, in the worst case, they can be fired. The actual cost of closing that hospital will thus depend whether we are able to re-employ them or not. 

Another complication which could arise in the executive restructuring has to do with the actual physical structure of the hospital network. For example, even if the proposed solution includes cutting of 100 hospital beds and diverting the same demand for healthcare services to less costly healthcare services (e.g. ambulatory services, day hospital, RSA, home care), it is not given that we could re-invest the current physical infrastructures (e.g. buildings, patients rooms, medical instrumentation) in the new services at zero costs. Generally speaking, there can be setup costs that we would need to consider. 

By having a look at the real allocation of beds among each clinical division of the various hospitals, one could consider to closedown those divisions with an improper ratio between personnel resources (doctors and nurses) and working beds, while adding the same (or a different) number of beds to a similar division within the same healthcare district or (if possible) within the same hospital. Indeed according to D.M. 13 Sep 1988 \footnote{\vv Determnazione degli standard del personal ospedaliero.", art. 3.} a Nephrology unit with 20 hospital beds should run with 6 doctors and 16 nurses but for any additional 20 beds unit only 3 doctors and 16 nurses are needed. By closing down small divisions and merging their beds to larger divisions, we could safe money.

It goes without saying that the evaluation of those costs strongly depends on the real executive implementation of the re-engineering plan. 
The aim of this paper is not the proposal of an executive but of a \bb{preliminary plan}. A preliminary plan is the first step toward the implementation of the most advanced executive plan. It gives us the flavour of the general feasibility of the project without going into much detail. Additional information relating to infrastructures, personnel and organization of the Apulian hospital network would allow to extend the preliminary proposal into an executive one.

\subsubsection*{Changing epidemiology}
The size of the hospital network has been calculated on current, i.e. historical, healthcare needs of the Apulian population. Clearly, if in the future the DRG distribution would be the same as the one reported for 2008, we are confident that the system has been properly sized. 
On the other hand, should the epidemiology of the Apulian population change, it is not guaranteed that the planned distribution of hospital beds would be the best one any more: it will then be not sure that future population healthcare needs would be satisfyied within the LEAs guaranteed by the Italian NHS.
Various scenarios are possible.\\

The proposed restructuring sees a relevent cut among acute ordinary beds in favour of day hospital and ambulatory services. However, not all diseases could be appropriately treated by day hospital or ambulatory services. If the morbidity rates increase affects diseases that need acute ordinary hospitalization, the system could turn out to be not capable of delivering the appropriate number of acute ordinary hospitalizations. This can translate into an inefficiency of the RHS and an increase of the overall expenditure, i.e. deficit could start accruing again. A new restructuring would then be needed.
On the other hand, if the morbidity spike affects those diseases that are appropriately managed through DH or ambulatory cares, chances are that our system can symply accomodate the higher demand by increasing the bed utilization rate. In this scenario, the hospital network will not need significant restructuring.\\

It goes without saying that forecasting epidemiologic trends is hardly feasible. Based on current events one can forecast with a certain degree of uncertainty what will be the trend in the next future (3-5 years).
For example there is increasing evidence of population ageing \cite{Ageing}. Population ageing can be foreseen to change epidemiology by shifting from acute to chronic diseases. Other particular events, such as flu and crisis can affect epidemiology.\\

By turning back to what discussed in the Introduction, a relevant problem to tackle would be to assess how will the economic crisis change current epidemiology among Apulian population.
Current economic crisis is one of the main event of the last 5 years and unfortunately the scale and depth of its consequences are not understood yet. This is the true question that one needs to answer in order to make sure the planned hospital network will continue being a good one in the next 3-5 years. One can make a starting assumption by looking at epidemiologic data of past crisis. Indeed, unemployment-related diseases are foreseen to play a big role.
Current DRG data can be tweaked in order to simulate different scenarios. This type of analysis - that we shall not provide within this paper - can allow to estimate a tolerance for the calculated allocation of hospital beds. The planned re-engineering could hence be improved.

However, epidemiologic studies have shown that crisis-related effects on healthcare take 3-5 years to realize. Current financial crisis started in 2008. In Italy, the unemployment rate, which is foreseen as one of the main risk factors for some crisis-related pathologies, rose one year later, in 2009. To some extent, Puglia is less exposed to the effects of the economic crisis given that most employed people work in public services. However, if on one hand public services hardly closedown, on the other hand due to specific measures hiring in the public sector is becoming increasingly difficult.  The bad economic consequences in terms of increasing unemployment rate are hence expected to be seen as long as the increasing demand for jobs of young generations will accrue without being met.

\cleardoublepage
\bibliographystyle{JHEP}

\begin{thebibliography}{99}
\bibitem{SDOdef}
SDO stands for the Italian \vv Scheda di Dimissione Ospedaliera" and is the final report that each NHS hospital must send to its Region when discharging a patient. The SDO is a fundamental management tool since it allows to monitor the performance of the hospitals and infer relevant information on financial costs.
\end{thebibliography}


\end{document}